\documentclass{usenixsubmit}
\pdfoutput=1
\usepackage{fullpage,epsfig,psfig,url,amsthm,amsmath,amsfonts,graphicx,myfig,pifont,color,cite,algorithm,algorithmic}
\usepackage{cite,fancyvrb,multirow}

\definecolor{LightGray}{gray}{.80}

\newcommand{\eat}[1]{}

\renewcommand{\footnotesize}{\small}

\makeatletter
{\catcode`\/\active\catcode`\_\active\catcode`\.\active
\gdef\URLslash{\futurelet\next\@@URLslash}%
\gdef\@@URLslash{\ifx\next\URLslash\char`\/\else\slash\fi}%
\gdef\URLdot{\char`\.\penalty\exhyphenpenalty}%
\gdef\URLprepare{\catcode`\/\active\catcode`\_\active\catcode`\.\active
        \let/\URLslash\let.\URLdot\def~{\char`\~}\def_{\char`\_}}}%
\def\URL{\bgroup\URLprepare\realURL}%
\def\realURL#1{\tt #1\egroup}%

\newsavebox{\sbrack}
\newsavebox{\mbrack}
\newsavebox{\lbrck}
\newsavebox{\mlbrck}

\date{}
\title{A Data Capsule Framework For Web Services: Providing Flexible
  Data Access Control To Users}
\author{
Jayanthkumar Kannan\\
UC Berkeley
\and
Petros Maniatis\\
Intel Labs Berkeley
\and
Byung-Gon Chun\\
Intel Labs Berkeley
}

\global\def\@maketitle{%
  \newpage
  \begin{center}%
  \let \footnote \thanks
\expandafter\ifx\csname acmdescription\endcsname\relax
  \null
\else
  {\setbox0\hbox{\vbox{%
\begin{flushright}%
  \begin{tabular}[t]{r@{}}%
    \acmdescription
    \\ \noalign{\vskip0.25in}%
  \end{tabular}%
\end{flushright}%
\null}}\ht0=0pt\dp0=0pt\box0}%
\fi
    \vskip -1.2em%
    {\LARGE \@title \par}%
    \vskip 2em%
    {\large
      \lineskip .5em%
      \begin{tabular}[t]{c}%
        \@author
      \end{tabular}\par}%
    \vskip 1em%
    {\large \@date}%
  \end{center}%
  \par
  \vskip 1em}
\newcommand{\eg}{{e.g., }}

\newcommand{\etal}{{et al.}}
\newcommand{\beq}{\begin{equation}}
\newcommand{\eeq}{\end{equation}}
\newcommand{\benq}{\begin{eqnarray}}
\newcommand{\eenq}{\end{eqnarray}}

\def\ged{\hbox{${\vcenter{\vbox{
        \hrule height 0.4pt\hbox{\vrule width 0.4pt height 6pt
        \kern5pt\vrule width 0.4pt}\hrule height 0.4pt}}}$}}

\newcommand{\bi}{
\begin{itemize}
\vspace{-0.05in}
\setlength{\itemsep}{-0.05 in}
}
\newcommand{\ei}{\end{itemize}}

\newcommand{\be}{
\begin{enumerate}
\vspace{-0.05in}
\setlength{\itemsep}{-0.05 in}
}
\newcommand{\ee}{\end{enumerate}}

\def\compactify{\itemsep=0pt \topsep=0pt \partopsep=0pt \parsep=0pt}
 \let\latexusecounter=\usecounter

\begin{document}
 
\maketitle

\vspace{-0.35in}
\begin{abstract}
  This paper introduces the notion of a secure data capsule, which
  refers to an encapsulation of sensitive user information (such as a
  credit card number) along with code that implements an interface
  suitable for the use of such information (such as charging for
  purchases) by a service (such as an online merchant). In our capsule
  framework, users provide their data in the form of such capsules to
  web services rather than raw data.  Capsules can be deployed in a
  variety of ways, either on a trusted third party or the user's own
  computer or at the service itself, through the use of a variety of
  hardware or software modules, such as a virtual machine monitor or
  trusted platform module: the only requirement is that the deployment
  mechanism must ensure that the user's data is only accessed via the
  interface sanctioned by the user. The framework further allows an
  user to specify policies regarding which services or machines may
  host her capsule, what parties are allowed to access the interface,
  and with what parameters. The combination of interface restrictions
  and policy control lets us bound the impact of an attacker who
  compromises the service to gain access to the user's capsule or a
  malicious insider at the service itself.
\vspace{-0.15in}
\end{abstract}

\section{Introduction} \label{sec:intro}

Internet users today typically entrust web services with diverse data,
ranging in complexity from simple credit card numbers, email
addresses, and authentication credentials, to datasets as complex as
stock trading strategies, web queries, movie ratings, and purchasing
histories. They do so with certain \textit{expectations} of
\textit{``what''} their data will be used for, \textit{``who''} it
will be \textit{shared} with, and \textit{``what''} part of it will be
\textit{shared}. These expectations are often violated in practice;
there are $400$ reported incidents of data loss from web services in
$2009$ per the Dataloss database~\cite{datalossdb}, each of which
exposed an average of half a million customer records outside the
service hosting those records. 

Such data exposure incidents can be broadly categorized into two
classes: \emph{external} and \emph{internal}. External violations
occur when an Internet attacker exploits software vulnerabilities at a
web service to steal sensitive user data, \eg credit card numbers.
Software and configuration complexity are often to blame here; the web
service does not benefit from these and, indeed, loses reputation as a
result. Internal violations occur when a malicious insider within a
web service, or even the service operator itself, abuses the
possession of sensitive user data beyond what the user signed up for,
\eg by selling customer marketing data to other companies. Both
external and internal violations are frequent: $65\%$ of the
afore-mentioned data exposure incidents fall in the external category,
while about $30\%$ fall in the internal category (the other $5\%$ do
not have a specified cause).

One option for a security-conscious user who wishes to limit the
impact of these threats is to avoid housing her sensitive data (say,
credit card number) at the service's site altogether. She can instead
host it close to herself: on her own machine. She then insists that
the service access this data over the network using a purpose-specific
interface. For instance, a credit card number need never be seen by a
merchant; the merchant site simply needs to request for a certain
charge to be made to the user's credit card number and receive an
authorization code to verify the transfer. This interface-based
approach assumes that the service \textit{refactors} their existing
applications (\eg shopping service) to work with such an
\textit{interface}. However, this \textit{client-side deployment}
option requires the user to have her machine online whenever the
service needs to access it (a problem with recurring credit
card charges, for instance) and incur bandwidth costs (which may be
significant in some cases, such as data analytics).

Alternatively, the user can choose to host her data a bit further away
from herself; she can enlist a ``data service'', an entity
adminstratively decoupled from the ``application service'' which
requires access to her data. We refer to such a remote data service as
a {\it trusted third party} (TTP). For instance, Amazon Payment is
such a TTP that encapsulates a user's credit card number with support
for restrictions such as limit on maximum charge. This option does
delegate the bandwidth consumption and reliability issues to the TTP;
however, for high-bandwidth interactions, the TTP would require
\textit{non-trivial provisioning cost} on the user. Further, the
\textit{flexibility} afforded to the user in terms of policy is
limited by what the data service provider offers.

\begin{figure}[t]
\begin{center}
\leavevmode
\mbox{\psfig{file=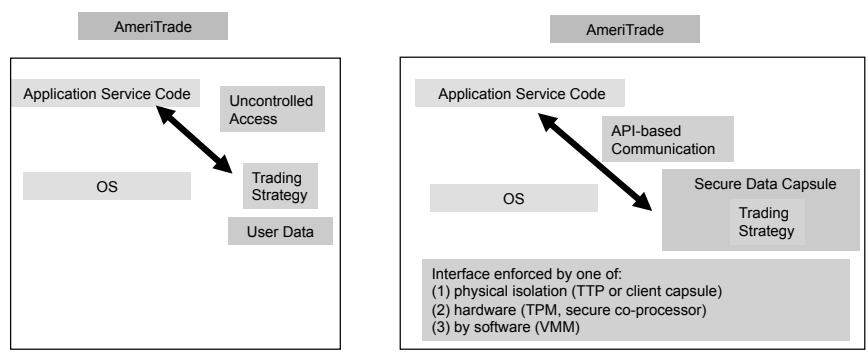,height=4in}}
\vspace{-1.85in}
\caption{ Trading Strategy Execution At AmeriTrade: (a) Current Model
  (b) The SDaC Framework}
\label{fig:intro}
\vspace{-0.35in}
\end{center}
\end{figure}  

Given the trade-offs of provisioning cost, performance, and level of
protection against threats provided by the options of service-side
data hosting, TTP-based hosting, and client-side hosting, the
motivation for a flexible framework that lets a user pick a suitable
deployment based on her needs is clear. In this work, we generalize
the concept of \textit{interface-based access approach} to
re-architecting web services that supports such {\it flexible
  deployment}. We built a capsule framework around this notion of
interface-based access allowing the freedom to enforce the interface
by any number of means; administrative isolation (as in a trusted
third party solution) or by software (such as a virtual
machinemonitor) or hardware module (such as a trusted platform
module).


Before elucidating the four main design principles that ground our
capsule framework, we present an instantiation of our framework in
Figure~\ref{fig:intro}. In the first option shown, the user (a
daytrader) reveals her trading strategy to AmeriTrade (an online
brokerage service) so that AmeriTrade can execute fast trades on his
behalf; however, the service is free to access the strategy as it sees
fit and may exploit it to its own advantages. In the second option,
AmeriTrade's view of the data is contrained by the interface; this
interface only allows AmeriTrade to report the current price of a
tradable instrument, and the only information revealed is the trade
the user wishes to make.


Our first design principle is that user's data should be accessed only
via a \textit{simple, narrow interface} that conforms to the principle
of least privilege. For instance, the only use a service has for a
credit card number is to charge it; access to the number itself is not
required. This let a user impose her own expectation of ``what'' is
done with her data, rather than rely on the service to enforce it. Our
main notion is that of a \emph{secure data capsule} (SDaC), an
encapsulation of a specific kind of data (say, a credit card number)
with code that implements a well-defined and open interface
suitable for the data (say, charging by a merchant). This open
interface model allows other parties to provider suitable
implementations, which can then be used by users in depositing their
data with web services. We demonstrate broad applicability of such an
interface-based model with examples from four significant application
classes: (1) a daytrading service in which large data volumes of stock
ticker data are parsed by complex private user queries to determine
automated trading actions. (2) a targeted advertising service in which
large volumes of per-user browsing history are mined. (3) a purchasing
application in which the capsule proxies requests for a credit card
number to a bank. (4) a provenance capsule that tracks changes to a
document with those who made them.

Our second principle is allow \textit{flexible interposition} of the
boundary between the capsule encapsulating a user's data and the web
service; such flexibility is crucial so that users can choose suitable
options to match their criticality, performance, cost, and threat
model requirements. Our framework supports, in addition to a trusted
third party model and client-side model, the use of trusted modules
(such as a virtual machine monitor) at application services to allow
{\it secure co-location} of the user's data at the application service
itself, whilest still guaranteeing interaction only via the interface.
Such a \textit{co-location based
  hosting model} addresses some of the limitations of the TTP Model:
it is applicable for high-bandwidth interactions, requires no
additional provisioning cost for the user, and can operate even under
disconnection. As an example, a trusted virtual machine monitor can be
used to logically isolate the capsule from the service. At a higher
per-invocation overhead, but for stronger isolation guarantees against
insider attacks, the capsule can be physically isolated from the
application service on distinct hardware collocated at the provider's
site. For even stronger isolation, while avoiding provisioning costs,
the capsule can be collocated with the user's client on a separate
virtual machine, but with high network overheads and reliability only
as high as that of the client machine.


The third design principle is allow for \textit{fine-grained and
  flexible user control} over how the interface to her data to her
exercised and where her capsule is hosted. This lets the user impose
her own requirements on ``what'' is done to her data and ``who'' her
data is shared with. Our framework includes a policy layer based on an
existing authorization language and mediates: (1) the parameters used
by the service in accessing her capsule's interface (2) transfers of
the user's capsule from one service to another. We note that this
policy layer is not part of the interface implementation since this
layer is meant to be controlled by users whereas the interface is
picked based on what functionality the service requires from the
service.

Our final design guideline is to allow a user direct control over {\it
  what part} of her data is shared; in particular, to allow
transformations of her data before any sharing. We support two
transformations currently; filtering and aggregation. Capsule
filtering enables a capsule at one service to spawn a derivative
version of itself at another service containing only a subset of the
user's original data. For example, a purchase history capsule at a
shopping site may spawn a version of itself with only music purchases
for the use of a music recommendation service. Although the same
semantics could be implemented via a policy extension (i.e., only
allowing the music service to read history entries pertaining to music
purchases), explicit filtering narrows the attack surface offered to
the music service, in the case where the isolation it offers were to
be violated. Capsule aggregation allows the merging of the data from
multiple capsules into an aggregate capsule that can be treated
monolithically by an application service. This is useful when services
simply require access to raw user data rather than
interface-restricted access; in such cases, one option is for users to
aggregate their data with others to gain privacy. For example, a user
and her friends may instruct their purchasing history capsules to
merge before interfacing with a recommendation service, to provide a
degree of anonymity for individual purchases across all friends
included.

\begin{table*}[t]
\begin{scriptsize}
\renewcommand{\arraystretch}{1.15}
\caption{Comparison of various deployment models}
\vspace{-0.1in}
\label{tab:intro_comparison}
\begin{center}
\begin{tabular}{|l|l|l|l|l|l|l|}
\hline\hline
Scenario & Flexibility of Policy & Performance & Provisioning
   & TCB & External & Internal \\
& & & Cost Borne By & & Threat & Threat \\
\hline\hline 
Raw Data & Limited by
app service & No network required & None & Unknown TCB (OS, apps) & No
& No \\
at app service & & & & & &\\
\hline
TTP Capsule & Limited by
data service & Limited by network & User & Trusted data
service & Yes & Yes  \\
\hline
Client Capsule & User can run own code  &
Limited by network & User & Open-source compact
TCB & Yes & Yes \\ 
\hline 
Co-located Capsule & User can pick own code & No network required &
Service & Open-source compact TCB & Yes & Yes\footnote{secure
  co-processor required} \\
\hline\hline
\end{tabular}
\end{center}
\vspace{-0.3in}
\end{scriptsize}
\end{table*} 

Table~\ref{tab:intro_comparison} summarizes the various options of
hosting the data in our capsule framework: the first is the current
case where data is stored at the application service, whereas the
other options are all supported by the capsule framework with the
consequent trade-offs.  We note that the co-located capsule can help
defend against both internal and external threats with the support of
a secure co-processor, and only against external threats if a virtual
machine monitor or trusted platform module is available. The security
guarantee provided by our framework against these threats is that the
combination of interface restrictions and policy control is an
upper-bound on the impact of an attacker. 

We have implemented a prototype SDaC framework that supports three
deployment models: TTP, client-side, and co-location using the Xen
hypervisor~\cite{xen_url} to isolate the data capsule from the
service. Our prototype implementation can only defend against internal
attacks in the client-side or the TTP model. Our co-location
implementation can only defend against external attacks, and {\it not}
internal attacks, since our implementation does not support a secure
co-processor. However, we believe it would be simple to extend our
implementation to support co-location based on TPMs and secure
co-processors, based on standard techniques (\eg
Flicker~\cite{flicker:eurosys08}).

We evaluate the SDaC approach along four axes: broad application
applicability, flexibility increase, performance, and TCB reduction.
We demonstrate broad applicability by refactoring exemplars from four
significant application classes. We demonstrate flexibility and
performance gains by evaluating the performance of each application in
a different deployment scenario, further showing that the optimal
deployment for each application differs, and significantly outperforms
other possible deployments. For instance, the co-location option
consumes no network bandwidth, as opposed to the TTP and client-side
capsules, which requires over $1$ Gbps network bandwidth per user in
the stock trading example. The Xen-based co-located capsule does incur
non-negligible access overhead (around $1$ ms) and storage costs
(around $120$ KB; though this can be amortized across users). However,
given the value of high-stake data, such as financial and health
information, and the fact that the cost of data exposure was estimated
by a recent study~\cite{ponemon:url} to be over $200$ dollars per
customer record, this overhead may be a acceptable price to pay for
controls over data exposure. The trustworthiness of our framework is
dependent on the SDaC implementation's correctness. As compared to a
complex and proprietary application service, the SDaC is typically
simple and can be divulged openly allowing for early discovery of
vulnerabilities. Our SDaCs do not require any complex OS services; our
design offloads the device drivers, network stack, etc to an untrusted
entity.

While we acknowledge that the concept of encapsulation is a well-known
one, we view as our main contribution a general architectural solution
to the problem of data access control in web services based on the
encapsulation principle, as opposed to custom solutions suited for
credit cards or targeted advertising. Our work is also related to
service composition approaches, but is specialized to providing
improved security, rather than improved functionality. 

Our capsule framework's applicability to a particular service is
limited by three main factors. First, capsules are not useful where
the interface required by the service involves {\it complex} function
calls (which precludes a simple capsule implementation) or when it
{\it reveals} the data directly (once the data is revealed directly to
an untrusted service, information flow control
mechanisms~\cite{jif:jsac03,fabric:sosp09,rifle:micro04,xbook:ssym09,asbestos:sosp05,dstar:nsdi08,resin:sosp09}
are required; data access control only controls release not
propagation). Second, our work also presupposes that a particular
interface is chosen by the service and vetted by the broader
community. It is a non-goal of our work to identify such interfaces
automatically for a service, or to prove that a given interface
provides a given level of privacy or other guarantees. For the various
web services we consider in this work, we suggest and implement
possible interfaces. Third, we assume that the service provider will
be willing to undergo a moderate amount of application refactoring,
possibly executing SDaC code they did not implement. We believe this
is realistic given that web services like Facebook already run
third-party applications as part of their business model.

The rest of the paper is structured as follows.
Section~\ref{sec:problem} defines our problem statement, while
Sections~\ref{sec:arch} and~\ref{sec:design} present an architectural
and design overview respectively of our capsule framework. We discuss
our implementation in Section~\ref{sec:implementation} and evaluate
its performance in Section~\ref{sec:evaluation} and its security in
Section~\ref{sec:secanalysis}. We then present related
work~\ref{sec:related} and then conclude in
Section~\ref{sec:conclusion}.

\section{Problem Statement}  \label{sec:problem}

The object of our capsule framework is to allow users {\it secure and
  flexible control} over {\it who} does {\it what} to their data, {\it
  when} they do it, and {\it where} their data is stored. Thus, our
goal is to enable flexible data access control. The {\it what} is
specified by the interface used to access the data, and is based on
{\it least privilege}; user data is exposed to the service only to the
extent required to accomplish the desired functionality. A
user-specified policy allows fine-grained control over the {\it
  what}, {\it who}, {\it when}, and {\it where} aspects of operations
performed on \textit{what} part of her data.

Two broad requirements for our capsule framework are: (1) Generality:
The framework must be broadly applicable across several kinds of web
services. (2) Flexible Hosting: The design should allow a variety of
hosting options depending on the trust assumption; co-location or TTP
or client-based.  The requirements of data-specific interfaces and
flexible hosting in our framework distinguish us from the two closely
related papers in literatures (Wilhelm's
thesis~\cite{mobile_privacy:thesis} and
Iliev~\etal~\cite{client_privacy:secpriv}); we discuss related work in
more detail in Section~\ref{sec:related}. We now flesh out our
security goals and assumptions
in more detail. \\

\noindent \textbf{Security Goal:} Our security goal is to ensure that
the policies specified by the user over her data are never violated.
In particular, any access to her data must be a sequence of legal
invocations of the interface sanctioned by the user that conforms with
her policy, and the only data exposed to the service is the output of
such a legal sequence of legitimate interface invocations. Note that
we only aim to circumscribe the influence of an adversary by the
limits set by the policy; the adversary will be able to exercise her
influence to the maximum allowed by the policy. For instance, if the
budget in a CCN capsule is specified as $100$ dollars, an adversary
will be able to exhaust this quota. However, the adversary will
neither be able to extract the CCN from the capsule, nor can she go
over $100$ dollars. There are two kinds of adversaries that we wish to
achieve this goal against.

\textit{Adversaries with software-only access:} This models an
Internet attacker external to the service who can exploit
vulnerabilites in the service software stack (above and including the
OS; we will assume the VMM, if any, is secure). Such adversaries are
responsible for the external category of attacks.  

\textit{Adversaries with physical access:} This models a malicious
agent at the service site who has physical access to the machine
hosting sensitive data; for instance, she may be able to monitor the
memory bus (say). This agent may be an admin or the organization
itself, who can launch internal attacks.

We present statistics from the DataLoss database~\cite{datalossdb} to
quantify the impact of these adversaries. Over the last decade,
adversaries with software-only access have been responsible for $65\%$
of data exposure incidents; there have been $394$ such incidents
caused due to hackers with an average of over $1$ million sensitive
user records exposed per incident. During the same period, adversaries
with physical access have caused $10\%$ of the incidents; there have
been $170$ such incidents with an average of over $340,000$ records
exposed per incident. We note that these two classes of adversaries
directly account for $75\%$ of the incidents in the database; of the
remaining $25\%$ incidents, the $21\%$ for which the cause is known
are due to accidental exposure due to inside elements. This category
includes incidents such as stolen/lost/disposed
computer/laptop/disk/drive/tapes and accidental mis-configuration of
software due to human error. For our purpose, the former kind of
incidents falls in the second class as the adversary has physical
access, while the second kind of incidents
falls in the software-only access case. \\\

\noindent \textbf{Security Assumptions:} Our framework can leverage
any of the following trust assumptions to ensure isolation between the
SDaC and the application service: \bi
\item Trusted Third Party (TTP): A TTP is an administratively
  different entity from the service that the user wishes to protect
  her data from. This trust can be based on a service-level agreement
  and the TTP would be paid for by the user. The option of hosting the
  capsule at the client's own computer is similar to the TTP option,
  since from the service's perspective, it interacts with the user's
  data over the network in both cases.

\item Trusted Hardware: A user can place her trust in a hardware
  module at the service site. Two kinds of hardware modules are
  currently available: Trusted Platform Module
  (TPM~\cite{tpmspec:tcg}) and Secure Coprocessor (\eg IBM
  4758~\cite{secure_coprocessor:computer01}). TPMs are now widely
  available in server machines, while secure co-processors, though
  preceding TPMs, are suitable for high-value data since they are
  considerably more expensive.

\item Trusted Software: A user can trust a virtual machine monitor
  (VMM, \eg Terra~\cite{terra:sigops03}) at the server site in order
  to achieve her goal. VMMs are now widely used in server
  environments, making this very viable. We envision two options here:
  attested VMs (where a TPM attests the execution of the VMM remotely
  to the client) and an un-attested VM (where the user trusts the
  service to invoke the VMM).

\ei 


Of these four trust assumptions, the TTP and the secure co-processor
models enable the capsule framework to defend against both threats
(internal and external), while TPMs and VMMs suffice to protect
against the threat of a software compromise. Once again, we note that
our prototype implementation \textit{cannot} defend against internal
attacks in the co-location model. \\










\noindent \textbf{Scope:} Our work currently make two assumptions.
First, we assume that a satisfactory interface has been arrived at for
a service of interest that provides the desired privacy to the user;
we do not attempt to verify that such an interface guarantees the
desired level of privacy, or to automate the process of re-factoring
existing service code. For the specific interfaces we present in this
work, we argue for their privacy based on informal arguments; we leave
formal verification for future work. Second, we assume that a given
capsule implementation carries out this interface correctly, and that
there are no side channels or bugs. For the capsules we discuss, we
strive to keep the interface simple and argue for the correctness of
the implementation based on the simplicity of the interface; proving
the correctness formally is a non-trivial research problem that we do
not aim to address in this paper. 


\section{Architecture} \label{sec:arch}

\begin{figure}[t]
\begin{center}
\leavevmode
\mbox{\psfig{file=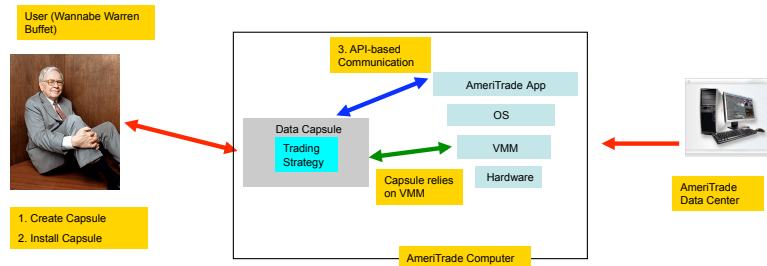,height=3.5in}}
\vspace{-1.85in}
\caption{The Capsule Architecture}
\label{fig:arch}
\vspace{-0.35in}
\end{center}
\end{figure}  

The main elements of our capsule architecture are presented in
Figure~\ref{fig:arch} that shows a user who capsulizes his trading
strategy at AmeriTrade's site; the capsule relies on a VMM for
security. The figure also shows the capsule lifecycle; the user
generates the capsule on her machine (say) and then transfers the
capsule to Amazon. This capsule then interacts with the data via a
prescribed interface. From hereon, we will refer to the user of a web
service as a {\it user} and to the web service as a {\it service}. In
the rest of this section, we will first examine the implications of
our framework in terms of how users and services interact with it. We
will then show that such an architecture is general and has several
applications; thus, our design principle of \textit{interface-based
  data access} is widely applicable.

\subsection{Implications to Users and Services}

A user, who wishes to use the capsule framework to protect her
sensitive data when stored at a service, picks a capsule
implementation from third party software companies (say a security
company such as Symantec) which exports an interface suitable for the
kind of data. Or, she may pick one from an open-source repository of
capsule implementations, or purchase from an online app-store. It is
even possible that the service itself offers a third-party audited
capsule implementation which the user may trust. For such a healthy
ecosystem of capsule implementations to exist, we envision that APIs
suitable for specific kinds of data will eventually be well-defined
for commonly used sensitive information such as trading strategies,
credit card numbers, email addresses, and web histories. Services that
require use of a specific kind of data can support such an API, which
would then be implemented by open-source reference implementations and
security companies. The development of such interfaces would have
another beneficial collateral effect; it would enable data portability
for users across services. Once the user picks a suitable
implementation, she then customizes it with policies that limit where
her data may be stored and who may invoke it. Our current
implementation requires users to use a declarative language for this
purpose; we envision that simpler user interfaces will be used for
this purpose. Once customization is done, the user initiates an
installation process by which the capsule is hosted (at a TTP or at
the service as desired) and the association between the service and
the capsule established.
 
From the perspective of the web service, the following changes need to
be made. First, they need to be willing to run third-party code in the
form of capsules. We believe this is reasonable since the
functionality of capsules is very limited; they can be executed within
a sandbox with simple policies (\eg allow network access to only the
payment gateway, allow no disk access). Further, a service can insist
that the capsule be signed from a set of well-known security
companies, similar to how applications are signed today. This gives
the service confidence in protecting its own code and data from the
capsule code. Given the success of Facebook applications, we believe
this model is reasonable. Second, a service may have to modify its
code to interact with the capsule via a programmatic interface,
instead of accessing raw data as they do today. This appears feasible
given the short life cycle of web service code (which are re-written
much more frequently than applications) and the fact that most web
services are architected so as to retrieve data from a remote machine
or via an application server. Modifying the Zen shopping cart
application~\cite{zen} to interact with the credit card capsule took
us only a single day. Third, the overhead of capsule storage and
invocation may be a hindering factor for adoption. The storage
overhead of the capsule seems reasonable especially if the capsules of
various users consists of code written by a small set of security
companies; thus, there would be considerable overlap in the code, and
the only additional storage required is for the user's data. We deal
with the invocation overhead in more detail in our evaluation section
(Section~\ref{sec:evaluation}), but for now, we note that the overhead
of invocation depends on the particular trust module in use. VMMs may
offer acceptable overhead since servers typically use highly optimized
virtualization already; several ways are known to optimize intra-VM
communication (\eg shared-memory based~\cite{xensocket:middle07}) and
to run in-line code securely without incurring inter-VM context switch
overhead (\eg using hardware paging features~\cite{secureinvm:ccs09}).
TPMs and secure co-processors incur significantly more overhead, and
may be feasible only for high-value data such as stock quotes and
health records. Late launch invocation in TPMs is still slow since
this feature is meant only for VM launches; however, in the future,
the trend is towards improving the performance of these devices
especially if they are used widely by web services.


\subsection{Notation}

We denote a capsule $C$ owned by user $U$ and resident on a machine
$M$ owned by principal $S$ as $C_U@[M,S]$. We denote the machine $M$
in the notation since whether or not a capsule $C$ can be hosted at
$M$ may depend on whether $M$ has a trusted hardware/software
module. Every capsule $C$ (we omit the service and the machine when
the exact capsule referred to is clear from the context) has a policy
database $P[C]$ specified by the user during creation. A capsule
$C_U@[MU,U]$ is created by the user $U$, for instance, on her own
machine $MU$. The user can then choose to host this capsule elsewhere,
say to a trusted third party $T$, by requesting a transfer
operation. The capsule $C_U@[M,U]$ then initiates a transfer operation
to the TTP $T$, at the end of which a capsule $C_U@[M,T]$ is now
hosted on the TTP $T$ ($M$ belongs to $T$). Alternatively, the user
can choose to host this capsule on the service's site itself by
relying on a VMM for security. In this case, the user's data is stored
at the capsule $C_U@[M,S]$ where $S$ denotes the service that owns
$M$. Once resident at $S$, the capsule $C_U@[M,S]$ can be accessed by
$S$; any invocations conforming with the user's policies (specified in
$P[C_U@[M,S]]$ are allowed by the capsule.

In order to model scenarios where one web service service can share
(with the user's consent) the user's data with another service, we
allow a service $S$ to share the user's data with a second service
$S'$. Alternatively, $S$ can proxy invocation requests from $S'$;
however, this makes $S$ liable for accesses that $S'$ initiates and
further, requires $S$ to incur overhead on behalf of $S'$. $S$ may not
desire such responsibilities and may prefer to transfer the capsule to
$S'$. To do so, $S$ requests the capsule $C_U@[M,S]$ to transfer to
$S'$. If the user's policy allows such a hosting, then the capsule
$C_U@[M,S]$ transfers itself to be hosted on $S'$ as $C_U@[M',S']$. We
note that, in order to support replication within the same service,
the service $S$ can ask that the capsule be hosted at another machine
$M'$ owned by the service $S$; this is treated as equivalent to an
across-service transfer. After the capsule has been installed, the
user also has the option of updating her data in various ways; we
support simple addition/redaction as well as allowing derivate
capsules that may have lesser data (by filtering) or may have
aggregate data from several users. Our subsequent
sections~\ref{sec:design_hosting},~\ref{sec:design_invocation},~\ref{sec:design_transformations}
deal with these three issues in more detail: hosting protocol for
tranfers, policies on invocation, and data transformations.



\subsection{Choice of Interface} \label{sec:arch_interface}

We now explain how the interface for a service is chosen, and will
illustrate the wide applicability of interface-based data access.
Denote by $F$ the service functionality that operates on the data; in
general, it is a function of $D_U$ (user's data) and $D_S$ (data
provided by the service). In the credit card scenario, $D_U$ is the
CCN number, $D_S$ is the merchant's account number, and $F$ contacts a
payment gateway and requests a transfer from CCN number $D_U$ to
merchant account number $D_S$.

When such functionality $F$ is provided by a capsule, it is refactored
as $F_C$ (implemented by the capsule), and $F_S$ (implemented by the
service) that operates on the output of $F_C$.
We will allow $F_C$ to operate on $D_U$ and $D_{SU}$ (which denotes
part of the service data $D_{S}$ which is revealed to the
capsule). The requirement is that the functionality remains the same
as before; thus, $F(D_U, D_S)$ is semantically equivalent to
$F_S(D_{S}, F_C( D_U, D_{SU}))$. The service reveals $F_C$ and
$D_{SU}$ (a subset of $D_S$) to the capsule, while the user reveals
$F_C(D_U,D_{SU})$ to the service. The final output of the function
$F(D_U,D_S)$ is typically exposed to the user, and in some cases, the
service as well. The order of invocation of $F_C$ and $F_D$ is
immaterial; we choose the formulation where $F_C$ acts first since we
are concerned with protecting the user. Using this notation, we
present some example classes of capsules; these classes are
distinguished based on a particular kind of application scenario. So,
it is useful to think of them as a starting set of usage idioms for
capsules. \\

\noindent \textbf{Query Idiom:} In this usage idiom, the server
provides a stream of incoming data, and the user would like to obtain
part of this stream or initiate actions based on this stream using a
filter without revealing either her filter or the matching items. In
this case, $D_S$ is this stream of incoming of data and $D_U$ is a
filter representing $U$'s interest in the data. The user's data is
thus like a continuous query in databases. For a capsule following the
query idiom, the following additional conditions hold: (a)
$F(D_{S,new},D_U)$ is the functionality that infers if the filter
$D_U$ matches a new data item $D_{S,new} \in D_S$. (b) $F$ is
well-known (c) $S$ reveals $D_{S}$ to the user since the user has paid
for this data already (\eg stock quotes) or because the stream $D_{S}$
is public. Condition (b) implies that the specs for $F$ can be
revealed publicly, enabling capsule implementers to support $F$.
Condition (c) implies that the service can share $D_S$ with the user's
capsule. These conditions ensure the capsule architecture is suitable
here.

Examples where this idiom applies include: (a) Stock Trading: Filter
is list of stock quotes that the user is interested and trading
algorithms that operate on ticker data. Stream of incoming data at the
server is high-rate financial ticker updates. The capsule
functionality makes trades based on ticker data. (b) Google Health:
Filter is list of medicines taken by user and list of diseases that
user has. Incoming stream at server consists of newly discovered
conflicts between certain diseases and certain medicines. A match
occurs if a new conflict involves user's medicines and diseases. (c)
Google News: Filter is user's list of interested keywords, while the
incoming stream is list of news articles. (d) Auction websites (\eg
Swoopo): Filter represents the bidding strategy of a user, whie
incoming stream is the current auction price and stream of bids from
other users.

A query capsule supports two interface calls: {\it Match} (used by the
service to update the capsule with a new data item in the stream) and
{\it RetrieveMatches} (invoked by the user to retrieve matching
items). Both are extremely simple to implement; {\it Match} adds the
new datum to a match buffer if a filter criterion matches, while {\it
  RetrieveMatches} authenticates the user and delivers matches. Care
is required in implementing these functions to avoid side channels
that expose the user's filter; we will address these in
Section~\ref{sec:implementation}. For now, we assume that the capsule
buffers up all matches within itself; whenever the user wishes to
receive updates, he {\it pulls} matching data items from the capsule.
This is to simplify the implementation; the {\it push} option is also
possible. Alternatively, the capsule may initiate some actions
automatically (such as make trades automatically) without any user
involvement. \\

\noindent \textbf{Analytics Idiom:} This idiom captures scenarios such
as targeted advertising, recommendation systems, where the service $S$
performs statistical analysis on the user's data, such as, web visit
logs, query logs, location trajectories, previous purchases. In this
idiom, the following hold: (a) $F(D_S,D_U)$ performs statistical
analysis on the user's data $D_U$ along with service data $D_S$. (b)
$F$ can either be proprietary or public (c) $D_S$ can either be
proprietary or public.This scenario is more complicated than our other
scenarios since both parties have private data: the user does not wish
to reveal $D_U$ while the server may not wish to reveal either $F$ of
$D_S$.

We will discuss the targeted ads scenario in detail. Here, $D_S$ is
the list of possible ads to display to the user which is not
proprietary. $F(D_U, D_S)$ selects one ad out of the list $D_S$ to
display and may be proprietary. We discuss the capsule interface
under two cases: when $F$ is well-known and when $F$ is proprietary.
There is one further choice to be made: whether $F(D_S,D_U)$, the ad
to be selected, is known to $S$. For simplicity, we assume that the
user only wishes to hide $D_U$ and not $F(D_S,D_U)$ since $D_U$ is
typically much more detailed than $F(D_S, D_U)$; it is possible to
extend this interface to hide the selected ad as well.

\textit{Well-known $F$:} This case is similar to the query capsule
case since both $F$ and $D_S$ are public. Thus, the re-factoring is:
$F_C = F$, $F_S = \phi$, and $D_{SU} = D_S$. $F_C$ is shown in
Algorithm~\ref{alg:targeted_ads_1} (we rely on
Adnostic~\cite{adnostic:ndss10} for the description of $F$). In this
description, $C$ denotes a category space representing various kinds
of user's interests such as ``Entertainment $\rightarrow$ Comics and
Animation $\rightarrow$ Cartoons'' , ``VideoGames $\rightarrow$ PS3''.
Interest categories are hierarchical; $\rightarrow$ represents a
sub-hierarchy. Google uses a 3-level hierarchy with around 600
categories overall. For simplicity of presentation, we ignore the
hierarchical structure within $C$ and consider it as a pre-specified
list of possible interest categories. $K$ denotes a keyword space
consisting of meta keywords that occur in webpages which convey
semantics; this keyword space is private to the capsule and is used in
the computation of $F$. The simplicity of this capsule is evident from
the conciseness of our pseudo-code.

\begin{algorithm}[t]
\caption{Select the ad to be shown based on user's history} 
\label{alg:targeted_ads_1} 
\begin{algorithmic}[1]
\STATE \textit{Function $F_C$ =
  ChooseAd(ListOfAds=$\{A_1,A_2,\cdots\}$, CategoryOfAds=$\{C_{i,c}\}$)
  where $C_{i,c}$ reflects the weight of ad $A_i$ in category $c$}
\REQUIRE \textit{Params:  $D_{SU} = \{A_1,A_2,\cdots,\}, \{C_{i,c}\}$
  \\ $D_U = \{W_1,W_2,\cdots\}$ ($W_i$ is set of websites
  visited by $U$), \\ $\{V_1,V_2,\cdots\}$ ($V_i$ is the number of visits to website $W_i$ by
  user $U$), \\
  $\{WK_{i,k}\}$ ($WK_{i,k}$ is the weight of webpage $i$ for
  keyword $k$; this is prepared by setting $W_{i,k}$ to one for words $k$ in the title bar
  and any meta keywords $k$ specified in the html source of the
  webpage, and then normalizing it), \\
 $DB = {DB_{k,c}}$ gives the weight for a keyword $k \in K$ to a
 category $c \in C$ (this database is public and can be prepared by
 crawling del.ici.ous bookmarks)}
\ENSURE \textit{returns Ad $A_i$ as selected ad}
\STATE $\forall~c~\in~C$, Compute $U_c$ (interest vector for user) as
$\frac{\sum_{i,k,c} ( V_i * WK_{i,k} * DB_{k,c})}{\sum_{i} V_i}$.
\STATE $\forall~A_i$, Compute $S_i$ (score assigned to Ad $A_i$) as
$\sum_{c \in C} C_{i,c} * U_c$.
\RETURN{$A_i$ st that $S_i$ is highest}
\end{algorithmic}
\end{algorithm} 

\textit{Proprietary $F$:} Google may choose to develop an algorithm
$F$ that depends on the user's browsing history, the ad's category,
the contents of the webpage that the ad is displayed on, and the
user's click through rates on ads previously shown to him. It is
difficult to anticipate what information may be used by Google, so in
such cases, we envision a capsule interface $F_C = F_C(D_U)$ (meaning
$D_{SU}=\phi$; that is, the service reveals nothing to the user), and
thus $F = F_S(D_S,F_C(D_U))$. One such interface $F_C$ is to return a
{\it perturbed} histogram of keywords that occured in web pages
visited by $U$; this perturbation is carried out using differential
privacy techniques~\cite{differentialprivacy:icalp06} which offers
provable privacy guarantees to the user. This interface gives
different privacy guarantees compared to the interface for public $F$;
in this case, the user gives up part of his privacy for the sake of
the complete privacy of Google's code. \\

\noindent \textbf{Proxy Idiom:} A capsule belongs to the proxy idiom
if the user has delegated some rights she has on a second service $S'$
to the capsule; the service $S$ uses the capsule to access $S'$ using
the user's right. We use the term proxy capsule since the capsule
which acts as a proxy between $S$ and $S'$. For such capsules, $D_U$
is typically a credential that is required to authenticate $U$ to
$S'$. For example, $D_U$ may be a credit card number, $S'$ is the
user's credit card company, and the capsule allows a merchant to
charge the user.

A proxy capsule offers two kinds of guarantees to the user. First, it
ensures that $D_U$ is not exposed to $S$. Second, it also constrains
the kind of operations that can be invoked under $U$'s authority on
$S'$. For a capsule following a proxy idiom, the following conditions
hold: (a) $F$ is a well-known function.  (b) $D_{S}$ is not
confidential information, so it can be exposed to the capsule. (c) $F$
requires interaction with a second service provider $S'$. Conditions
(a) and (b) imply that following factoring is acceptable to $S$: $F_C
= F$, $F_D = \phi$, and $D_{SU} = D_S$. Condition (c) implies that the
capsule functionality can be offloaded to the service provider $S'$ in
some cases, but this is not always possible if $S'$ does not support
the kind of policies that $U$ would like. Condition (c) also implies
that the capsule requires network access, typically over SSL.  We now
discuss payment capsules, one particular instantiation of a proxy
capsule, and then discuss other instantiations.

\begin{algorithm}[t]
\caption{Charge user's credit card number and tranfer requested amount 
to merchant's account } 
\label{alg:ccn_capsule} 
\begin{algorithmic}[1]
\STATE \textit{Function $F_C$ = Charge(Amt, MercAcct)}
\REQUIRE \textit{Params: $D_{SU} = (Amt, MercAcct)$, Amt $>$ 0, \\
 $D_U = CCN \text{(set by user)}$, \\
 CapsuleState: PaymentGateway (set to dns name of payment gateway such
 as ``gateway.authorize.net''), \\ PaymentGatewaySSLName(set to
 VeriSign-signed name of gateway such as ``Authorize.Net''), \\
 RequestTemplate (set to template HTTP POST request for charging)}   
\ENSURE \textit{RetVal $=$ ConfCode if successful, $= -1$ otherwise}
\STATE Request $=$ RequestTemplate
\STATE Request $=$ Replace(Request, ``Amount'', Amt)
\STATE Request $=$ Replace(Request, ``MerchantAccount'', MercAcct)
\STATE Response $=$ SubmitOverSSL(PaymentGateway,PaymentGatewaySSLName, Request)
\STATE ConfCode $=$ RegexpExtractBody(Response,/conf\_code=/)
\RETURN{ConfCode}
\end{algorithmic}
\end{algorithm} 

A \textbf{payment capsule} allows a service $S$ to charge user $U$ on
a recurring basis while allowing the user to impose policies on such
charging (\eg do not charge more than once a month) and protecting the
user's charging token (such as a CCN or a gift certificate or e-cash)
from the service. The interface $F_C$ for such a capsule and its
implementation is shown in Algorithm~\ref{alg:ccn_capsule}. $D_{SU}$
is passed into the function $F_C$ during invocation, and $D_U$ is
stored within the capsule permanently.  We use the notation {\it
  Capsule State} to indicate scratch state retained across
invocation. The notation {\it Require} and {\it Ensure} indicate the
pre-condition and the post-condition for the function
respectively. Lines $1-4$ construct the POST request to be sent. The
two functions in this capsule dispatch the request and parse the
reply: {\it SubmitOverSSL} and {\it RegexpExtractBody}. {\it
  RegexpExtractBody} is a simple regular expression matching function
used to extract the confirmation code from the server's response. {\it
  SubmitOverSSL} is responsible for contacting the gateway over SSL
and obtaining the server response to the constructed POST request. We
will explain in Section~\ref{sec:implementation} how this function can
be implemented by leveraging an {\it untrusted} network stack; for
now, we will note that it can be implemented with the aid of a trusted
cryptographic library and an untrusted network stack.

Two other instantiations of a proxy capsule are access capsules and
notification capsules; due to space constraints, our discussion below
is limited. An \textbf{access capsule} allows a service $S$ limited
access to the user's data stored at a second service $S'$, while
keeping the user's authentication token for $S'$ (\eg password) secret
from $S$ and imposing restrictions on accesses made from $S$ to
$S'$. For instance, an access capsule is used by a user $U$ to allow
FaceBook restricted access to her contacts list on Gmail, and hide the
rest of her Gmail data (emails etc) from Facebook. Other examples
include services like Mint that require access to the user's
up-to-date bank financial transactions at say, Bank of America. A
\textbf{notification capsule} allows a service $S$ to send messages to
a user $U$, while the user $U$ need not reveal anything about how and
where she can be reached, and can specify policies on message sending
(such as rate limiting). An example is a email capsule where $F$
implements sending the message via email to $U$, $D_U$ is $U's$ email
address, $D_S = \phi$, $S$ is an online merchant, $S'$ is $U's$ mail
server, and $F(D_S,D_U)$ returns a success code if the delivery was
succesful. \\

\noindent \textbf{Provenance Idiom:} In this idiom, the user's data
$D_U$ is modified by various parties that have access to it; the goal
is to ensure that each modification is tied to the party which
modified it. Thus, for this case, $F(D_U,D_S)$ updates the data $D_U$
and $D_S$ is the modifications made by the party to which the capsule
is forwarded. The security guarantee is thus an integrity requirement
rather than a privacy requirement. The various parties are willing to
reveal $F, D_U, D_S$ to each other; the security guarantee is that a
data modification $D_S$ is tied to the party $S$ that made the
modification.

We envision such a capsule being useful in a enterprise scenario for
collobarative editing.  A provenance capsule containing a text file
(say, a legal memo or a business plan) is forwarded by a user to a
group of people who will collaboratively edit the data; the data owner
wants to ensure that any changes to the data is tied to the person who
made the change. This usage is reminscent of a distributed version
control system with additional security requirements.

A provenance capsule supports two calls: \textit{ReadData} and {\it
  WriteData)}. Upon invocation, the capsule authenticates the caller
and returns the data (for {\it ReadData}) or updates it along with a
signature of the modification by the caller (for {\it WriteData}). The
implementation is simple: {\it WriteData()} appends the modification
to a write log along with the signature of the invoking principal as a
proof of provenance, {\it ReadData()} returns the current version of
the data. \\


\noindent \textbf{Summary:} The four idioms above capture a wide set
of scenarios; of these, the proxy capsule is the most complicated
since it requires network access. For these scenarios, the capsule
interface is simple and provides the required privacy guarantees to
the user; we believe that this simplicity bodes well for a correct
implementation. Capsules are not useful where the interface involves
{\it complex} function calls or when it is {\it leaky}. As an example
of the first, consider the interaction between Google Docs and a
user's document. While one can theoretically envision an interface
(\eg display the first page on the screen, delete character at line 20
and position 30), the interface would be so rich as to make it
difficult to characterize leaks and to give any confidence as to the
correctness of the capsule implementation. The interaction between
Facebook and a user's profile data is an illustration of the second
case. Facebook requires direct access to the user's data (\eg name,
address), and such an interface affords no privacy to the user; once
the capsule releases the data to a Facebook process, it is not
possible to ensure any policy constraints. These two scenarios are
better handled by information flow control (\eg
XBook\cite{xbook:ssym09}) or end-to-end encryption (\eg
NOYB~\cite{noyb:wosp08}).

\section{Design} \label{sec:design}

\begin{figure}[t]
\begin{center}
\leavevmode
\mbox{\psfig{file=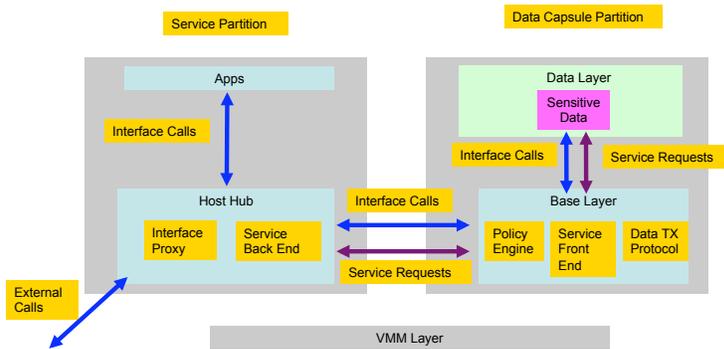,height=3.5in}}
\vspace{-1.4in}
\caption{Design Overview}
\label{fig:design_overview}
\vspace{-0.35in}
\end{center}
\end{figure}  

In this section, we discuss the design of the capsule architecture in
more detail. The previous section mainly focussed on our design
principle of interface-based access; the components of the base layer
we discuss in this section are based on the principles of (a) flexible
deployment with policy control. (b) policy control over invocation.
(c) policy control over data on which the invocation operates.

Figure~\ref{fig:design_overview} shows a capsule co-located at the
service site that uses a VMM as a trusted module to isolate itself
from service code. It shows a {\it host hub} that resides within the
service partition and a {\it base layer} in the capsule partition (we
use the term partition to imply the isolation between these
components). We assume the base layer is installed by the service
within the VM allocated for a user's capsule; during the installation
protocol, the user can verify that this is the case using attestation.
The host hub acts as the proxy for communication between the external
world and the capsule. The base layer includes generic functionality
(such as hosting protocol, policy checking) common to all capsules,
while the data layer implements interface calls specific to the data
item.

The figure also shows the three main components of the base layer: the
policy engine, the data transformation module, and the service front
end. The policy engine is used by the base layer to check whether an
invocation request or an hosting request should be allowed or not.
Regarding invocation requests, we are guided by the principle of
providing fine-grained flexible control, and regarding hosting
requirements, our goal is to to allow a range of flexible deployment
options. The data transformation module is responsible for
implementing a secure aggregation protocol (filtering is data
specific, and are implemented in the data layer). The service front
end is responsible for carrying out complex functions (such as network
access, disk access) which are out-sourced to the service back end in
the host hub which uses device drivers to provide this functionality
to the capsule (the concept of front end and back end is similar to
how Xen supports devices inside a guest VM). Such \textit{service
  requests} are made by the data layer and are proxied to the service
back end by the service front end.  This split architecture
significantly reduces the complexity of the capsule and our TCB; the
capsule only relies on the host hub for availability, not for
confidentiality or integrity guarantees on the data. The only code
inside the capsule correspond to the base layer functionality and data
layer functionality; typical OS components such as device drivers,
network stack, are not required.

The rest of the section presents the detailed design. We first discuss
the interaction between the host hub and the base layer
(Section~\ref{sec:design_interaction}). We then discuss the policy
engine in two parts: the hosting policy engine
(Section~\ref{sec:design_hosting}) and the invocation policy engine
(Section~\ref{sec:design_invocation}). We then finally discuss the
data transformation module (Section~\ref{sec:design_transformations}).

\subsection{Interaction between Host Hub and Base Layer} \label{sec:design_interaction}

The interaction between the host hub and the base layer is secured by
the base layer by leveraging the trust module. If the capsule is
hosted at a TTP or at the client, then the trust is guaranteed by the
physical isolation between the capsule and the service. If the capsule
is co-located at the service, then the base layer requires the
following functionality from the trust module: (a) isolation (b)
non-volatile storage. (c) random number generation (d) remote
attestation (optional). Isolation is needed to ensure that the service
code cannot interfere with the capsule and steal its data during
execution. Non-volatile storage is required to defend against replay
attacks. Otherwise, it is possible for an adversary to rollback the
capsule state to an earlier point in time in an undetectable fashion;
this would allow them to violate the security guarantee desired by the
user. Randomness is required in order for secure key and nonce
generation in the various protocols that the base layer implements.
Remote attestation is optional; it lets the trust module can prove to
the user that her data is indeed processed by the capsule. We will
later argue in Section~\ref{sec:secanalysis} that these properties
suffice for the security of our framework; for now, we note they are
provided by all the three trust modules we consider (VMM, TPM, secure
co-processor). We now introduce the host hub and the base layer.

The host hub has two main roles. The first is to serve as a proxy for
all communication to and from the capsule; such communication may be
to/from the user (during capsule installation) or to/from the service
code (during invocation). This design decision lets us decouple the
application on the service side from the details of capsule invocation
and is a modularity-based decision. The second role is to provide
services to the DC such as network access and persistent storage; this
allows the capsule to leverage functionality without having to
implement it, considerably simplifies the implementation. The host hub
runs within a un-trusted OS, but this does not violate our trust,
since the capsule only relies on the host hub for availability, not
safety: data is encrypted before any network communication or disk
storage via the host hub. Based on the various data layer capsules we
have implemented, we have identified a set of common services they
require: \bi \item Raw Network Access via the BSD Socket API: When a
data layer
  (such as a CCN capsule) requires network access, in order to avoid
  placing a complex TCP/IP network stack within the capsule, the base
  layer offloads the implementation to the host hub. For instance,
  when the data layer invokes the {\it connect} BSD socket call, the
  base layer marshals the parameters and conveys them to the host
  hub. The host hub invokes the {\it connect} call using its network
  stack (which need not be trusted) and returns a file descriptor (or
  error) to the base layer which conveys it to the base layer. The
  data layer can thus avail of the standard socket API.  We have also
  included the PolarSSL SSL/TLS library~\cite{polarssl:url} in the
  base layer and modified the socket calls to use the base layer's
  rather than using the standard system calls. We chose the PolarSSL
  library since it is a significant smaller code base ($12$K) compared
  to OpenSSL (over $200$K lines of code); this design decision can be
  revisited if required. Thus, data layers can use the SSL library as
  well; this is necessary to access any remote web service securely.
\item Time: The base layer can also request the host hub to retrieve
  the current time from a remote trusted NTP server using SSL; this is
  useful for supporting certain policies (for instance, expire the
  data after a certain date).
\item Persistent Data: The host hub also offers a simple block-based
  storage interface for reading and writing data persistently. None of
  our capsules use this feature so far (since they live within a VM,
  and are re-incarnated along with the VM image), but this
  functionality may be required when large data sets are involved.
  \ei

  The base layer handles all requests from the host hub and implements
  the hosting protocol and the invocation policy. It also interacts
  with the trusted module to ensure that the security properties
  hold. Thus, the data layer that implements the data-specific
  interface is decoupled from the security mechanisms, and is much
  easier to write. We now describe the components of the base layer.



  \subsection{Capsule Hosting} \label{sec:design_hosting}

In order to meet our requirement of allowing flexible data hosting
policies, our capsule framework allows user to specify a {\it hosting
  policy} which is then enforced by a {\it hosting protocol}. We first
discuss our hosting policy language and then discuss the hosting
protocol.

\noindent \textbf{Hosting Policy:} The requirement for our hosting
policy language is that it be expressive enough to accomodate typical
user requirements and be simple enough to be interpreted easily by the
framework. We arrived at a design that is a simplified version of
SecPAL~\cite{secpal:csf06}, an authorization policy language for
distributed systems. The simplifications were done so as to enable
easy interpretation. Our language is flexible enough to allow a user
to grant a service $S$ the right to host her data based on: (a)
white-list of services (b) trusted hardware/software modules available
on that service. (c) if a service $S'$ trusted by the user vouches for
the service $S$. This space of options is inspired from the typical
privacy options supported by websites. The flexibility to trust based
on the availability of a trusted modules is key; a user may chose to
trust any service that houses her data in a physically secure
co-processor, while it may only trust a select set of services to host
her data on a VMM or a TPM. We show an example below to illustrate the
flavor of our
language. \\\\
\colorbox{LightGray}{\parbox{1.0\textwidth}{ \texttt{Alice}
    \underline{says} \textit{CanHost(\texttt{M})} \underline{if}
    \textit{OwnsMachine(\texttt{Amazon},\texttt{M})},
    \textit{HasTPM(\texttt{M})} \\
    \texttt{Alice} \underline{says} \texttt{CA} \underline{can say}
    \textit{HasTPM(\texttt{X})} \\
    \texttt{Alice} \underline{says} \texttt{S} \underline{can say}
    \textit{OwnsMachine(\texttt{S},\texttt{M})} \\
    \texttt{Alice} \underline{says} \texttt{Amazon} \underline{can
      say} \textit{TrustedService(\texttt{S})}  \\
    \texttt{Alice} \underline{says} \texttt{Amazon} \underline{can
      say} \textit{CanHost(\texttt{M})} \underline{if}
    \textit{TrustedService(\texttt{S})},
    \textit{HasSecureCoprocessor(\texttt{M})},
    \textit{OwnsMachine(\texttt{S},\texttt{M})} \\
    \texttt{Alice} \underline{says} \texttt{CA} \underline{can say}
    \textit{HasSecureCoprocessor(\texttt{X})} }}

~\\We use a typewriter font to indicate principals like Alice (the
user; a principal's identity is established by a certificate binding
its name to a public key), an underlined font to indicate language
keywords, and italics for indicating predicates (such as {\it CanHost}
and {\it HasTPM}). The first rule says that \texttt{Alice} allows any
machine $M$ to host her capsule provided Amazon acknowledges a machine
as its own, and the machine has a TPM. The second rule says that
\texttt{Alice} trusts \texttt{CA} to certify the public keys of a
machine's TPM. The third rule allows any service to certify a machine
as its own. The fourth rule allows Amazon to recommend any service $S$
as a trusted service. The fourth rule indicates Alice allows any
machine $M$ to host her capsule if (a) Amazon vouches for such a
service $S$ (b) $S$ asserts that $M$ is its machine (c) the machine
$M$ has a secure co-processor (since Alice is delegating the decision
to Amazon, she would like a more secure option than a TPM, so her
policy asks for a secure co-processor). The last rule indicates
\texttt{Alice} delegates her decision of which machines have secure
co-processors to the certification authority \texttt{CA}. This example
illustrates all features of our policy language; the simplification
from SecPAL is that we do not support recursive delegation.

A hosting request received at $C_U@[M,S]$ has three parameters: the
machine $M'$ to which the hosting transfer is requested, the service
$S'$ which owns machine $M'$, and a set of assertions $P_{HR}$. The
set of assertions $P_{HR}$ are presented by the requesting principal
in support of its request; it may include delegation assertions (such
as ``\texttt{S} \underline{says} \texttt{S'} \textit{CanHost(X)}'') as
well as capability assertions (such as ``\texttt{CA} \underline{says}
\textit{HasTPM(M')}''). When such a request is received by
$C_U@[M,S]$, it is checked against the policy $P(C_U@[M,S])$. This
involves checking whether the fact ``\texttt{U} \underline{says}
\textit{CanHost(M')}'' is derivable from $P(C) \cup P_{HR}$. If it is,
then the hosting protocol is initiated.


 
\textbf{Hosting Protocol:} The hosting protocol is responsible for the
forwarding of a capsule from one party to another. We treat the
initial transfer of a capsule from the user's machine to a TTP or
service as a hosting transfer as well. Thus, capsule $C_U@[M_U,U]$
denotes the capsule initially created and resident on the user's
machine. The hosting protocol is based on the Diffie-Hellman
key-exchange protocol, and has three steps: \bi
\item Step 1: $M \rightarrow M'$: $K_C$, $N$
\item Step 2: $M' \rightarrow M$: $K_{C'}$, $Attestation(M',\text{BL},N,K_C,K_{C'})$
\item Step 3: $M \rightarrow M'$: $ \{ C_U@[M,S] \}_{DHK} $
\ei 

The only differences from the Diffie-Hellman protocol (involving the
keys $K_C, K_{C'}$ and the Diffie-Hellman secret $DHK$) are the use of
the nonce $N$ and the optional attestation that proves that the key
$K_{C'}$ was generated by the DC base layer code which, on input $N,
K_C$, produced output $K_{C'}$. This attestation is made along with an
attestation key $M'$ which is vouched off for by a certification
authority, guarantees freshness (since it is bound to $N$), and rules
out man-in-the-middle substitution on both the input and output (since
the attestation is bound to both $K_C$ and $K_{C'}$). At the end of
this three-message exchange, the base layer at $M'$ then de-serializes
the freshly transferred capsule $C'_{U}@[M',S']$ identified by the key
$K_{C'}$. At this point, the capsule $C'$ is operational; it shares
the same data as $C$ and is owned by the same user $U$.




\noindent \textbf{Consistency:} We note that when a user's data is
spread across several services in the form of several capsules, we
{\it do not} attempt to enforce any notion of consistency across them
{\it after} they have forked off; each of them is an independent
entity (much like the case today; if Amazon hands out a user's credit
card number to a third party, it does not necessarily update it
automatically on any changes by the user). Our architecture allows a
user or service to notify any descendant capsules and update them if
so desired, but this is not automatically done since it is not
possible to maintain both consistency and availability in the face of
partitions, and further, the overhead of synchronizing the multiple
copies would be high. However, {\it during} the process when one
capsule is forked off from another, we appropriately transfer policies
during the transfer so that the policies continue to hold.

\subsection{Capsule Invocation  Policy} \label{sec:design_invocation} 

An invocation policy allows the user to specify constraints on the
parameters to the capsule interface during invocations by various
principals. We support two kinds of constraints: stateless and
stateful.

Stateless constraints specify conditions that the argument to a single
invocation of a capsule interface must satisfy. For example, such a
constraint could be ``never charge more than 100 dollars in a single
invocation''. We support predicates based on comparison of numerical
quantities, along with any conjunction or disjunction
operations. Stateful constraints apply across several invocations; for
example, ``no more than 100 dollars during the lifetime of the
capsule''. We found such stateful constraints to be useful for
specifying cumulative constraints. It is very likely that users would
desire to specify a budget on their credit card number over a period
of time. An even more important use of a stateful constraints is in
specifying a query budget for differential privacy based
interfaces. 
Thus, it is crucial for users to be
able to specify a query budget that limits the number of times their
data is accessed; this requires maintaining state of the number of
queries served so far. Once again, we aim for the policy language
supporting both these kinds of invocation constraints to be simple;
this language is very similar to the hosting policy language and we
present an example for a
CCN capsule below. \\\\
\colorbox{LightGray}{\parbox{1.0\textwidth}{ \texttt{Alice}
    \underline{says} \textit{CanInvoke(Charge,\texttt{Amazon},A)}
    \underline{if} \textit{LessThan(A,100)}\\
    \texttt{Alice} \underline{says}
    \textit{CanInvoke(Charge,\texttt{DoubleClick},A)} \underline{if}
    \textit{$LessThan(A,Limit)$, Between(CurrentTime, `Jan 1st,
      2010'',``Jan 31st, 2010'')} \underline{state} $(Limit=50,Update(A))$ \\
    \texttt{Alice} \underline{says} \texttt{Amazon} \underline{can
      say} \textit{CanInvoke(Charge,\texttt{S},A)} \underline{if}
    \textit{$LessThan(A,Limit)$} \underline{state}
    $(Limit=150,Update(A))$ }}

~\\The first rule says gives the capability ``can invoke upto amount
A'' to Amazon. This allows Amazon to call the charging interface with
parameter $A$ and the constraint $A<100$ checks that the amount is
less than $100$. The second rule shows a stateful example; the
semantics of this rule is that DoubleClick is allowed to charge upto a
cumulative limit of $50$ during Jan 2010. To implement this policy, we
introduce a state variable called {\it Limit} which is set to $50$
initially by the user. The predicate {\it Update(Limit,A)} is a
stateful predicate that indicates if this rule is matched, then the
{\it Limit} should be updated with the amount $A$; thus, when a rule
is matched with a {\it state} keyword, it is removed from the policy
database (the database of assertions), the state variable (\eg {\it
  Limit}) is updated in the rule, and the new rule inserted into the
database. This usage idiom is similar to SecPAL's support for RBAC
dynamic sessions; we have added the {\it state} keyword to allow
retention of state between authorization requests. The alternative is
to move this state outside the SecPAL policy, and house it within the
capsule functionality; we make the design choice of making it explicit
in the policy itself to ensure that the policy implementation is not
split across SecPAL and the capsule implementation of the
interface. Though this choice implies that the policy database is
updated over time, there are no undesirable consequences with allowing
such dynamic changes. The third rule is very similar to the second
rule; the interesting thing to note here is that this rule is matched
for any principal to which Amazon has bestowed invocation rights. This
means that the limit is enforced across all those invocations; this is
exactly the kind of behavior a user would
expect.  

\textbf{Transfer of Invocation Policies:} We now discuss how this
invocation protocol interacts with the hosting protocol discussed
earlier. During a hosting protocol initiated from $C_U@[M,S]$ to
$C'_U@[M',S']$, $C$ should ensure that $C'$ has suitable policy
assertions $P(C')$ so that the user's policy specified in $P(C)$ is
not violated. First, any policy statements $P_{HR}$ specified by $S'$
during the hosting request need to be added to $P(C')$ to record the
fact that this newly derived capsule operates under that context.
Second, any stateful policies need to be specially handled. For
example, consider the third rule in the invocation policy in
Section~\ref{sec:design_invocation} which requires that the total
budget across all third parties that are vouched for by Amazon is
$100$ dollars. If this constraint is to hold across both $C'$ and any
future $C''$ that might be derived from $C$, then one option is to use
$C$ as a common point which ensures that this constraint is violated.
However, this requires any transferred capsule $C'$ to communicate
over the network with $C$ upon invocations. Instead, in order to allow
disconnected operation, we use the concept of
exo-leasing~\cite{exoleasing:mdw08}.

In the exo-leasing concept, the constraint (\eg budget) is broken up
into sub-constraints (\eg sub-budgets) so that if the sub-constraints
are enforced, the parent constraint is automatically enforced. In the
capsule context, {\it decomposable constraints} (such as budgets,
number of queries answered so far) from $AC(C)$ are split into two
sub-constraints; the original constraint in $AC(C)$ is updated with
one sub-constraint, and the second sub-constraint is added to
$AC(C')$.  In our example, the assertion ``\texttt{Alice}
\underline{says} \texttt{X} \textit{can invoke Charge A State Limit
  100} \underline{if} \texttt{Amazon} \underline{says} \textit{I vouch
  for X}, \textit{ [ $ ( A < Limit ) $ , Update (Limit,A) ] }'' would
be transformed into ``\texttt{Alice} \underline{says} \texttt{X}
\textit{can invoke Charge A State Limit 75} \underline{if}
\texttt{Amazon} \underline{says} \textit{I vouch for X}, \textit{ [ $
  ( A < Limit ) $ , Update (Limit,A) ] }'' and ``\texttt{Alice}
\underline{says} \texttt{X} \textit{can invoke Charge A State Limit
  25} \underline{if} \texttt{Amazon} \underline{says} \textit{I vouch
  for X}, \textit{ [ $ ( A < Limit ) $ , Update (Limit,A) ] }''.


\subsection{Capsule Data Transformation} \label{sec:design_transformations}

In exploring the use of capsules in various application scenarios, we
found it useful to allow users to perform transformations on their
data post-installation, especially when the data is of an aggregate
nature (\eg web history). Providing users control over the
transformations on their data is distinctly different from providing
control over invocation; the latter controls operations invoked over
the data, and the former controls the data itself. We refer to a
capsule whose data is derived from a set of existing capsules by some
transformation as a derivative capsule. We support two data
transformations: filtering and aggregation.



\noindent \textbf{Filtering:} A derivative capsule obtained by
filtering has a subset of the data of the originating capsule; for
instance, only the web history in the last six months, instead of the
entire year. A capsule that supports such tranformations on its data
exports an interface call for this purpose; this is invoked alongside
a hosting protocol request so that the forwarded capsule contains a
subset of the originating capsule.

\noindent \textbf{Aggregation:} This allows the merging of raw data
from mutually trusting users of a service, so that the service can use
the aggregated raw data, while the users still obtain some privacy
guarantees due to aggregation. We refer to this scenario as ``data
crowds'' (inspired by the Crowds anonymity
system~\cite{crowds}). Examples of usage include ads clicking behavior
of users or online product purchase history. In the first, Google
(say) may use the aggregate ad click behavioral history to fine-tune
their targeted advertising algorithms. In the second, the latter could
be used by Amazon to build statistical models for recommendation
systems. 

To enable this, a user $U$ instructs her capsule $C_U@[M,S]$ to
aggregate her data with a set of capsules $C_{U'}[M',S]$ where $U'$ is
a set of users that she trusts (this trust is necessary; otherwise, if
her data was merged with data belonging to fake users, the privacy
guarantees would be much poorer). These set of users $U'$ form a data
crowd. We envision that a user $U$ can discover such a large enough
set of such users $U'$ by mining her social networks (for instance).

During capsule installation, each member $U$ of the crowd $C$ confides
a key $K_C$ shared among all members in the crowd to their
capsule. During installation, a user $U \in C$ also notifies the
service of her willingness to be aggregated in a data crowd identified
by $H(K_A)$. $S$ can then identify the set of capsules $C_A$ belonging
to that particular data crowd using $H(K_A)$ as an identifier. The
service can, at this point, then determine any kind of aggregation
strategy. For instance, in order to reduce network overhead, it could
request all the capsules stored on a rack in its data center to
aggregate with each other first, and then ask the resultant capsule to
aggregate with the resultant capsules from other racks. Note that the
aggregation protocol does reveal the size of individual user data;
such side-channels can be avoided by padding data if desired. 

From the capsule's perspective, aggregation is a pair-wise operation;
a capsule $C_U$ is requested to merge with $C_{U'}$, and these
mergings are appropritately staged by the service. During the
aggregation operation of $C_U$ with $C_{U'}$, capsule $C_U$ simply
encrypts its sensitive data using the shared data $K_A$ and hands it
off to $U'$ along with its owner's key. During this aggregation, the
resultant capsule also maintains a list of all capsules merged into
the aggregate so far; capsules are identified by the public key of the
owner sent during installation. This list is required so as to prevent
duplicate aggregation; such duplicate aggregation can reduce the
privacy guarantees. Once the count of source capsules in an aggregated
capsule exceeds the user-specified constraint $C_{min}$, the aggregate
capsule can then reveal its data to the service $S$. This scheme
places the bulk of the aggregation functionality upon the service;
this is ideal since it gives the service freedom to optimize the
aggregation, while at the same time requiring the capsule to only
implement a simple pair-wise aggregation function.



\section{Implementation} \label{sec:implementation} 

Our implementation supports three deployment models: TTP, client-side,
and Xen-based co-location. In the future, we plan to support TPMs and
secure co-processors using standard implementation techniques using as
late launch (\eg Flicker~\cite{flicker:eurosys08}). We implement four
capsule instances atop this framework, one per usage idiom: a stock
trading capsule, a credit card capsule, a targeted ads capsule, and a
provenance capsule. We describe some of the details of our framework
followed by a description by each capsule.

\subsection{Capsule Framework} \label{sec:impl_framework}

\begin{table*}[t]
\begin{scriptsize}
\renewcommand{\arraystretch}{1.15}
\caption{Capsule Implementation: LOC Estimates of Modules}
\label{tab:tcb}
\begin{center}
\begin{tabular}{|l|l|l|}
\hline\hline 
Software Module & LOC & Functionality \\
\hline\hline
Base Layer & 6K &  Implements installation, invocation, and policy
checking. \\
\hline
Data Layer & Stock ($340$), Ads ($341$)  & Implements functionality
specific to data  \\
& CCN ($353$), Provenance ($156$) & \\
\hline
PolarSSL & 12K & Light-weight cryptographic and SSL library used for
embedded systems.  \\ 
\hline
XenSocket & 1K & Fast inter-VM communication using shared memory and
event notification hyper-calls. \\ 
\hline 
Trousers (optional) & 10K & Library for TPM functions. Can be removed
from TCB by accessing TPM directly.  \\ 
 \hline\hline
\end{tabular}
\end{center}
\end{scriptsize}
\end{table*}

For TTP deployment, the network serves as the isolation barrier
between the capsule and the service.  For co-location deployment, we
used the Xen virtual machine~\cite{xen_url} as the isolation
mechanism; the web service code and the data capsule are run inside
different virtual machines thus providing protection against software
compromise. We ported and extended XenSocket~\cite{xensocket:middle07}
to our setup (Linux 2.6.29-2 / Xen 3.4-1) in order to provide fast
two-way communication between the web service VM and the capsule VM
using shared memory and event channels. 

The base layer implements the Diffie-Hellman based installation
protocol and invocation protocols. It implements policy checking by
converting policies to DataLog clauses, and answers query by a simple
top-down clause resolution algorithm (described in
SecPAL~\cite{secpal:csf06}; we could not use their implementation as
such since it required .NET libraries). We use the PolarSSL library
for embedded systems for light-weight cryptographic functionality. We
use a TPM, if available, to verify that a remote machine is running
Xen using the Trousers library; this TCB can be reduced by invoking
the TPM directly rather than via the library. Note that we only use a
TPM to verify the execution of Xen; we still use Xen as the isolation
mechanism. The host hub uses the \textit{libevent} event library (the
host hub is not part of the TCB). We use 2048-bit RSA public keys (for
identifing capsule instances, data owners, services), 256-bit AES/DES
keys (used internally when encrypting data using public keys), and
HMAC/SHA-2 hash functions (for signatures).

We now estimate the TCB of our Xen based capsule. This consists of the
Xen VMM, Dom0 kernel, and our capsule implementation.
Table~\ref{tab:tcb} shows the LOC estimates of the various modules in
our capsule implementation. Our capsule implementation currently has
two dependencies which we plan to remove in the future. First, our
capsule VM boots up the capsule on top of Linux 2.6.29-2; however, the
capsule code does not utilize any Linux functionality, and can be
ported to run directly atop Xen or MiniOS (a bare-bones OS distributed
with Xen). Second, the capsule uses the memory allocation
functionality by \textit{glibc} and some functionality from the STL
library; we plan to include a custom memory allocation module to
substitute \textit{glibc} and to hand-implement custom data structures
to avoid this dependence.

We also plan to incorporate mechanisms focussed on improving VMM-based
security (such as removing Dom0 functionality from the TCB by using
disaggregation techniques~\cite{xendisaggregation:sigops08}); such
mechanisms are orthogonal to our framework. We have yet to incorporate
several performance optimizations discussed in literature for
improving performance while providing isolation. One technique that is
pertinent is from Sharif~\etal~\cite{secureinvm:ccs09}.
\cite{secureinvm:ccs09} allows us to run the capsule within the same
VM as the service, while still guaranteeing code and data isolation;
this is achieved by having the VMM use hardware paging functionality.
\cite{secureinvm:ccs09} demonstrates an order-of-magnitude gain in
performance using this technique, which we hope to implement in our
framework. We now discuss the implementation of the base layer's
internals followed by the sample capsules we implemented.


\subsection{Capsule Hosting} \label{sec:impl_hosting}

We currently use a policy language which supports three base types:
principal names, strings, and integers. An argument in a predicate can
either be a constant or a variable, and is annotated with its type,
\eg VPRINCIPAL(XP) denotes a variable XP which stands for a
principal. A policy in our language is represent as a 5-tuple $(P, F,
CF, C, S)$ where $P$ is name of the principal issuing this statement
(such as \textit{CPRINCIPAL(ALICE)} which denotes a constant ALICE of
type PRINCIPAL), $F$ is the statement made by $P$ (\eg \textit{FACT\{~
  PRED [~INVOKE, ~VPRINCIPAL(XP), ~VNUM(AMT), ~VSTR(ACCOUNT) ~] ~\}}
), $CF$ and $C$ is a set of conditional facts and constraints which
are required to be satisifed if $F$ is to be true, and $S$ records any
state variables.  In order to verify whether a particular statement is
derivable from a set of policy statement, we translate all statements
to DataLog clauses, and then use a simple top-down resolution
algorithm (described in SecPAL~\cite{secpal:csf06}). We defer a
detailed description of our policy language and resolution algorithm
to the technical report due to space constraints.

\subsection{Invocation Policies} \label{sec:impl_invocation}

An invocation request has four parameters: $S'$, the service making
the request (not necessarily the service $S$ hosting the capsule), the
request type $R$ (\eg ``Charge''), the arguments to the request $A$
(\eg ``10'' dollars), and a set of supporting assertions $P_{IR}$. To
service such a query, the base layer of $C$ checks to see if the fact
``\texttt{U} \underline{says} \texttt{S'} \textit{can invoke $R$
  $A$}'' or the fact ``\texttt{U} \underline{says} \texttt{S'}
\textit{can invoke $R$ $A$ State $S$}'' is derivable from $P(C) \cup
P_{IR}$. If there is no proof for such a query, the request is
denied. If there is a proof, the base layer retrieves any policy
specification $ac_{S}$ in the proof with a {\it State} keyword. For
all such assertions $ac_{S}$, the corresponding state $S$ is also
extracted. The invocation is then dispatched by the base layer to the
data layer; after the successful completetion of the request, the
state $S$ is updated to $S'$ as per the {\it Update} rules. At this
point, the old specification $ac_{S}$ is removed from $P(C)$, and a
new specification $ac'_{S'}$ is inserted into $C_U@[S,M]$. This
ensures that the state is updated as required. To avoid race condition
issues, for now, the base layer waits for the invocation to complete
before serving any other requests. Since the database of policies may
change from one invocation to the next, we do not cache any
intermediate facts that are derived duing one invocation for
subsequent ones; thus, we avoid any inconsistency issues.

\subsection{Data Transformations} \label{sec:impl_transformations}


Filtering is easy to implement; a capsule $C$ supports an interface
call {\it Filter} which takes a filtering criterion as an argument,
and produces a capsule $C'$ per the filter. Filtering can be invoked
in conjunction with a hosting transfer; alternatively, the policy can
insist that it should only be invoked by the owner of the capsule. For
the latter case, every capsule maintains the public key of its owner
$K_C$ (which is sent during the installation and hosting protocols),
and authenticates requests for data transformation using this key.

Aggregation is a more expensive operation since it involves
coordination of $U$'s capsule with multiple other capsules. To enable
secure aggregation, our aggregation protocol operates in two stages;
the first stage is performed off-line by a set of users $U_A$ who have
decided to cooperatively perform aggregation of their data stored on a
common service $S$, the second is the aggregation itself performed at
the service site.

During the setup phase, a set of users $U_A$ establish a shared key
$K_A$ amongst themselves; a user $U$ can securely discover such a set
of willing users $U_A$ and establish a key $K_A$ using her social
network. All the user's implementations support an aggregation call in
their interface. At the end of this phase, every user $U \in U_A$
notifies her own capsule $C_U@[M,S]$ of $K_A, A_{min}$. Here $A_{min}$
represents the minimum number of capsules that should be aggregated
before the data is released; we return to this later. Note that this
notification step can be performed during the installation
itself. Further, this set of users $U_A$ can be the same across
multiple services. Thus, a user need only establish this $U_A, K_A$
once; this set can be re-used across multiple services.

The second phase performs the actual aggregation itself, and is
performed with the coordination of the service. During installation,
a user $U \in U_A$ notifies the service of her willingness to be
aggregated in a data crowd identified by $H(K_A)$. $S$ can then
identify the set of capsules $C_A$ belonging to that particular data
crowd using $H(K_A)$ as an identifier. The service can, at this point,
then determine any kind of aggregation strategy. For instance, in
order to reduce network overhead, it could request all the capsules
stored on a rack in its data center to aggregate with each other
first, and then ask the resultant capsule to aggregate with the
resultant capsules from other racks.

From the capsule's perspective, aggregation is a pair-wise operation;
a capsule $C_U$ is requested to merge with $C_{U'}$, and these
mergings are appropritately staged by the service. During the
aggregation operation of $C_U$ with $C_{U'}$, capsule $U$ simply
encrypts its sensitive data using the shared data $K_A$ and hands it
off to $U'$ along with its owner's key. Thus, the shared key $K_A$
serves as a security token; only another capsule belonging to the same
crowd will be able to decrypt the data. During this aggregation, the
resultant capsule also maintains a list of all capsules merged into
the aggregate so far; capsules are identified by the public key of the
owner sent during installation. This list is required so as to prevent
duplicate aggregation; such duplicate aggregation can reduce the
privacy guarantees. Once the count of source capsules in an aggregated
capsule exceeds the user-specified constraint $A_{min}$, the aggregate
capsule can then reveal its data to the service $S$. This scheme
places the bulk of the aggregation functionality upon the service;
this is ideal since it gives the service freedom to optimize the
aggregation, while at the same time requiring the capsule to only
implement a simple pair-wise aggregation function.

\subsection{Sample Capsules}

The broad guideline behind our sample capsule implementations is to
model the essence of the application scenario in realistic enough
fashion so that experimental comparative analysis of different
deployment options is feasible.

\noindent \textbf{Stock Trading Capsule:} We model our stock capsule
functionality after the automatic trading functionality in a popular
day trading software, Sierra Chart~\cite{sierrachart:url}, which is
representative of the complex mechanisms used by financial traders.
We chose one specific feature of Sierra Chart for illustration: make
trades automatically when an incoming stream of ticker data matches
certain conditions. Our implementation follows the the Sierra Chart
technical manual's description~\cite{sierrachart:url}.

This capsule belongs to the query idiom (Section~\ref{sec:arch_interface})
and exports a single function call of the form
\textit{TickerEvent (SYMBOL, PRICE) } and returns a \textit{
  ORDER(``NONE'' /` `BUY'' / ''SELL'', SYMBOL, QUANTITY) } indicating
whether the capsule wishes to make a trade and of what quantity. The
capsule allows the user to specify two conditions: a BUY (ENTRY)
condition and a BUY(EXIT) condition. A condition is an arbitrarily
nested boolean predicate with operations like AND, OR, and NOT, and
base predicates that consist of numerical variables, comparison
operations ($>, <, =$), and numerical constants. The numerical
variables include: LP (the latest price of stock), MA (moving average
of price), POS (position: amount of stock purchased by the software),
and POSAV (average price of stock purchased already). As an example, a
BUY(ENTRY) condition could be \textit{$LP>MA$} and a BUY(EXIT)
condition could be $OR(AND(>0,LP<=POSAV-1), AND(POS>0, LP>=POSAV+2))$.
This means that the system purchases upto $Q$ (a user-chosen
parameter) everytime the last price exceeds the moving average, and
this stock is sold off either when: (a) last price is less than the
average price the BUY Entry was filled at minus 1 full point. (b) the
last price is greater than the price the Buy Entry was filled at plus
2 full points.
Our implementation of boolean predicate matching is straight-forward;
we do not apply any query optimization techniques like common
sub-expression matching from the database community. Thus, our
performance results are only for comparative purposes.

Sierra Chart currently executes at the user's desktop on a feed of
incoming stock symbols (similar to the TTP case). In the Xen case, the
user capsulizes her BUY(ENTRY) and BUY(EXIT) conditions at her broker.
The only information revealed by the capsule is any trades it makes;
the strategy is itself secret. It is not possible to hide the trades
from the broker, thus any leaks of the strategy through the trades
cannot be avoided; we assume that this leak is tolerable. We expect
that the re-factoring of the application service at the broker site
would not require significant effort since the interface is very
simple; since we do not have access to any broker site code, we can
only estimate this effort.

\noindent \textbf{Credit Card Capsule:} We implemented a credit card
capsule that interacts with the \texttt{authorize.net} gateway to
implement charges, and modified an open-source shopping cart
application (Zen) to interact with the capsule to implement charging.
As discussed in Section~\ref{sec:arch_interface}, this capsule
supports a single interface call, \textit{Charge(Amt, MercAcct)}, and
returns a confirmation code to the service. This capsule illustrates
the flexibility of the framework with respect to invocation policies
and matches the proxy idiom. This functionality is offered by some
banks already; we only implement and evaluate this capsule to measure
the cost of our framework in emulating this functionality in a more
general fashion. We implemented both stateless policies (allow only
invocations for less than $X$ dollars) and stateful policies (maintain
a per-month budget of $X$ dollars).

\noindent \textbf{Targeted Advertising Capsule:} We implemented two
capsules for targeted advertising both of which belong to the
analytics idiom (Section~\ref{sec:arch_interface}). They both store
the user's web browsing history and are updated periodically (say, on
a daily basis). The first capsule is used by a service to serve
targeted ads when a user visits a specific web page. The second is
used by the service to build long-term models related to the affinity
of users with specific profiles to specific advertisements. These
models are used, for instance, by Google, to fine-tune their targeted
advertising methods. Both capsules support a simple interface call
that is used by the data owner to update her web browsing history; we
now detail the other interface calls exported by these capsules below.

\textit{Serving targeted ads:} This capsule supports two possible
interfaces: \textit{ChooseAd(List of Ads, Ad Categories)} and
\textit{GetInterestVector()}. In the first case, the capsule selects
the ad to be displayed by using a procedure similar to those used in
web services today (per the description in
Section~\ref{alg:targeted_ads_1}). In the second, the capsule extracts
the keyword vector of the user from her browsing history, and then
computes the true normalized interest vector $U = \{ U_c \}$ based on
the user's data $(D_U = \{W_1,W_2,\cdots\})$ where $\{W_i\}$ is the
user's web visit history. Here, $0 \leq U_c \leq 1$ reflects the
interest of the user in category $c$ and $\Sigma_{c} U_c = 1$. The
capsule then obfuscates this vector per a generic differential privacy
prescription. We obfuscate each element $U_c$ of this vector $U$
individually, and then re-normalize the vector $U$. The $L_1$
sensitivity of this function, denoted by $\Delta(U_c)$, is $1/V$ where
$V$ is the total number of visits by the user to various websites
since if the user visits one particular different website, the count
of the number of visits in one specific category $c$ can at most
increase or decrease by $1$. We insist that the user visit a minimum
number of websites $V_{min}$ before the capsule would start serving
ads. Thus, the sensitivity $\Delta(U_C)$ of this function is
upper-bounded by $1/V_{min}$. The obfuscated value $U'_C$ of $U_c$ is
chosen per the distribution function $Pr[U'_c = a] \propto exp (
\frac{ - \parallel (U_c - a) \parallel }{\sigma} )$ where \textit{exp}
is the scaled symmetric exponential distribution function and $\sigma$
is the variance of the distribution. This function guarantees
$\epsilon$-differential privacy for $\epsilon = \Delta(U_c)/\sigma
\geq 1 / (V_{min}\sigma)$.

This capsule also makes use of a stateful policy in the differential
privacy case to record the number of queries made so far. The amount
of information leaked by interactive increases linearly with the
number of queries made~\cite{differentialprivacy:icalp06}; thus it is
necessary for users to specify a private budget $Q_{max}$ and the
framework ensures that this budget is enforced by using stateful
policies (in a manner similar to the credit card budget). As discussed
earlier, these two calls provide different privacy guarantees. The
first leaks no information; the selected ad is sent to the user's
computer where a client-side component displays the ad. In the second,
one implication of the $\epsilon$-differential privacy guarantee is
that even if the service knows all but one of the websites visited by
the user, the ability of the service to deduce the one private website
is limited by $\epsilon$. In particular, the distributions of the
output for various possibilities of that one private website are
point-wise $\epsilon$ close. We choose a typical value $\epsilon =
0.01$. We 


\textit{Building long-term models:} The second capsule allows the
service access to the websites visited by the user and the particular
advertisements clicked by the user when visiting a website (thus this
data is more detailed than the data used for targeted
advertising). This information is used to refine the service's
targeting algorithms. For this purpose, it suffices for the service to
obtain aggregate data from a large population of users; we can thus
leverage our aggregation functionality. Individual user data is
represented as $D_U = \{ W_i \}, \{ (P_i, A_i, C_i) \} $ where $\{ W_i
\}$ represents all websites visited by the user, and $\{ (P_i, A_i,
C_i) \}$ represents how the user responded to ads displayed on a
publishing websites $P_i$ that participates in the advertising service
(such as Google AdSense). $C_i$ is $0/1$ depending on whether the user
clicked on ad $A_i$ shown to her when visiting website $P_i$. This
information is aggregated across all users by our aggregation protocol
and delivered to the service. To do so, the users distribute a shared
secret key amongst themselves which is then used as an authentication
token amongst the capsules to authenticate themselves.  Currently, we
simply aggregate this information and return it to the service; if
desired, differential privacy techniques can be applied here as well.

In the future, we plan to merge these capsules with a client-side
browser plugin to build a full-fledged system so that the plugin can
automatically update the server-side capsule without any manual
intervention from the user. This plugin would also interact with the
capsule to obtain and display the selected ad in the case of the {\it
  ChooseAd} interface. If it is necessary for the server to maintain
ad impression counts (for charging advertisers), then information
about the selected ad can be sent after aggregation or via a broker to
preserve anonymity; our system can borrow such functionality from
PrivAd~\cite{privad:hotnets09} and
Adnostic~\cite{adnostic:ndss10}. Alternatively, our aggregation
functionality may be used for this to obtain aggregate ad impressions
as well.


\noindent \textbf{Provenance Capsule:} The final capsule we
implemented belongs to the provenance idiom
(Section~\ref{sec:arch_interface}) and models a editable
document. It supports three calls: \textit{Get()},
\textit{Insert(Position,String)}, and \textit{Delete(Position,
  NumChars}. These calls suffice for any sequence of edits; we assume
an editor program propagates any edits made by colloborators as calls
to this interface so that the capsule is altered. \textit{Get()}
returns the current version of the document, while {\it Insert()} and
{\it Delete()} can be used by, say, an editor, to convey changes made
by a user to the document. The capsule retains a log of all
invocations, thus binding each modification to the invoker. This
illustrates that the capsule framework can also be used to provide
properties such as accountablity in addition to privacy.

\noindent \textbf{Other Possible Capsules:} There are also services
where the some private data may be required when the service is
disconnected from the Internet during invocation of the capsule so
that a trusted third party solution is not feasible. One example is a
service like CommonSense~\cite{commonsense:url} where sensors are worn
by users that record information about the user. This information may
be processed by proprietary algorithms (for example, to monitor blood
pressure); such algorithms have to be implemented at the sensor itself
because the user may not be connected to the Internet at all times. In
the future, we hope to implement this capsule.

\section{Evaluation} \label{sec:evaluation}

We evaluate our framework based on four capsule instances (one per
usage idiom): stock trading capsule, credit card capsule, targeted ads
capsule, and provenance capsule. We omit the provenance capsule's
results since they reflect similar trends as for the other
capsules. For the three capsules we evaluated, we present three
scenarios: (A) the base case where the service interacts with the raw
data with no access guarantees. (B) the TTP case, where the service
interacts over the network with the user's capsule hosted at a
TTP. Since a TTP deployment is equivalent in terms of performance in
most respects to a client-side capsule, we discuss both of them
together, and mention any differences specifically. (C) the co-hosting
case, where the service interacts with the data capsule housed in a
Xen VM.

Our goal in this section is two-fold. Our first goal is to to evaluate
\textit{the cost of privacy} as the performance impact of the
client-side / TTP deployment and co-location deployment options, as
compared to the base case. Our metric of success here is hard to
quantify in terms of absolute numbers; we only wish to show that this
cost does not render our framework impractical. The second goal is to
compare \textit{the various deployment options} amongst themselves to
determine the ideal deployment for each scenario.

We evaluate the performance for these scenarios along three
dimensions: the setup cost (when the data or data capsule is
transferred to the service), the invocation cost (when the service
accesses the user's data or data capsule), data transformation costs
(the cost of operations such as aggregation). The term {\it cost}
includes latency, network bandwidth consumption, and storage cost. Of
these, the invocation cost is borne every time the user's data is
accessed, and is thus the primary metric of comparison. The setup cost
is incurred during the initial and subsequent transfers of the user's
data, while the transformation cost, thought not as frequent as
invocation, may be significant if the transformation involves data
belonging to large numbers of users (such as aggregation).

Our test server is a $2.67$ GHz quad core Intel Xeon processor with
$5$ GB memory on which Xen, the service code, and the capsule code
run. A desktop (Intel Pentium 4 $1.8$ GHz processor, $1$ GB memory),
on the same LAN as the server, served as a TTP (or client). The
bandwidth between the server and this desktop was $10$ Gbps and the
round-trip about $0.2$ milli-seconds. To simulate a wide-area network
between the client/TTP and the service, we used
DummyNet~\cite{dummynet:url} which artificially delays packets by a
configurable parameter; by default, the round-trip is $10$ ms (typical
wide-area latencies from home users to well-provisioned services like
Google are about $30$ ms; we chose $10$ ms under the conservative
assumption that a TTP might have faster connectivity than average).


\subsection{Framework Benchmarks}

We first evaluate our framework independent of any specific kind of
data. For these measurements, we used a dummy data layer that does not
do any processing; it simply accepts a invocation payload of a
specific size and returns a response of specific size to help in
benchmarking.

\begin{table*}[t]
\begin{scriptsize}
\renewcommand{\arraystretch}{1.15}
\caption{Invocation Cost ($1$ KB sensitive data, $1$ KB invocation
  request, $1$ KB invocation response )}
\label{tab:invocation_cost}
\begin{center}
\begin{tabular}{|l|l|l|l|l|}
\hline\hline 
Scenario & Latency (micro-seconds)  &  Bandwidth &
In-Memory Storage &
On-Disk Storage \\
& (Mode, $50\%$-ile range, $95\%$-ile range)  & (bytes) & (KB) & (KB) \\
\hline\hline
Base Case & $471.5~[379,496]~[373,532]$ & NA & 1 (data) & 1 (data) \\ 
\hline
TTP & $24796~[24718,24867]~[24525,25033]$ & $3336$  & 120 (capsule) + 1 (data) & 120 (capsule) + 1 (data) \\ 
\hline 
Xen Capsule & $1237~[1097,4203]~[942,4617]$ & NA & 120 (capsule) + 1 (data) & 120 (capsule) + 1 (data) \\
 \hline\hline
\end{tabular}
\end{center}
\end{scriptsize}
\vspace{-0.3in}
\end{table*}


\textbf{Invocation Cost:} The invocation costs for a fixed payload
size of 1 KB are shown in Table~\ref{tab:invocation_cost} (the
invocation request and response are both $1$KB). For the base case, we
assume that the service is architected as an server-side application
that interfaces with a database process to access the user's data. We
assume that the database process runs on the same machine as the
server-side application (ignoring any transfer costs) and that the
user's data is in memory (ignoring any costs of retrieving the data
from disk). For the TTP case, the web service interacts with the
capsule over the network. For the Xen capsule, the web service process
accesses the data via the host hub module which invokes the capsule
housed in a different VM instance. We assume that the capsule VM is
housed on the same machine as the service VM (similar to the base
case). We measured the latency as the time required for the process
storing the user's data to respond to an invocation by the web
service; for this benchmark, we fixed the message sizes for the
invocation request and response to be the same in all three scenarios.
The bandwidth consumed is measured using \textit{tcpdump} and includes
packets in both directions as well as TCP headers. The in-memory
storage cost and on-disk storage cost is measured directly. These
results are reported in the table; all results are reported over $100$
runs. For latency, we report the median, the $50\%$ confidence
interval, and the $95\%$ confidence interval, since it is somewhat
variable; we report only the median for bandwidth since we found it to
be much less variable.

Though the latency via Xen is about $800$ micro-seconds worse compared
to the base case, it is still significantly lower compared to the TTP
case. We found considerable variation ranging from $900$ micro-seconds
to $4000$ micro-seconds with a median of $1237$ micro-seconds; we
believe this is related to the Xen scheduling algorithm which may
delay the execution of the capsule VM. We plan to use techniques based
on hardware virtualization~\cite{secureinvm:ccs09} that allow for
order-of-magnitude reductions in overhead by executing the capsule
inside the service VM while still guaranteeing VM-like isolation
(since capsules require no access to service VM state except via the
interface, the ``semantic gap'' issue does not arise).

In the TTP case, the latency is primarily due to two round-trips (one
for exchange of TCP SYN and SYN ACKs, and another for the invocation).
The TTP option consumes $3.3$ KB worth of network traffic per
invocation, as opposed to the Xen co-location option and the base case
where the invocation is local. The invocation request and the response
are $1$ KB each: a constant overhead of about $1$ KB is incurred since
our protocol sends invocation requests and responses over RPC/HTTP
using \textit{libevent}. In both the TTP and Xen case, the capsule
code has to be stored in memory and disk; in our current un-optimized
implementation, this code is about $120$ KB. The LOC of the capsule in
this case is only $111$ LOC; our current compilation technique bundles
the glibc library as well which leads to this additional overhead. In
practice, we believe that a single capsule implementation will be
shared by several users (for instance, supplied by Symantec); thus a
single copy of the capsule code need not be maintained for every 
user. 

We now present a component-wise breakdown of the cost of a invocation
request; recall that a request is made by the service to the host hub
which then relays it to the Xen capsule. In this experiment, the
average invocation latency was $1399$ micro-seconds from the service's
perspective. Of this, $101$ micro-seconds is due to host-hub
processing, and $768$ micro-seconds due to processing at the Xen
capsule; thus, $530$ micro-seconds was due to communication overhead
(between the service and host hub, and between the host hub and the
capsule). Within the Xen capsule, $165$ micro-seconds was spent in
verifying the invoker's signature and $109$ micro-seconds for policy
verification; the remaining $494$ micro-seconds was due to the
processing of the $1$ KB invocation payload and response. In
conclusion, communication overhead is about $35\%$ of the total
latency; the rest is primarily due to processing within the Xen
capsule (which may be slower since it is run in a guest VM).

\begin{figure}
	\vspace{-0.1in}
        \centering
        \mbox{
                \begin{tabular}{c}
                        \psfig{file=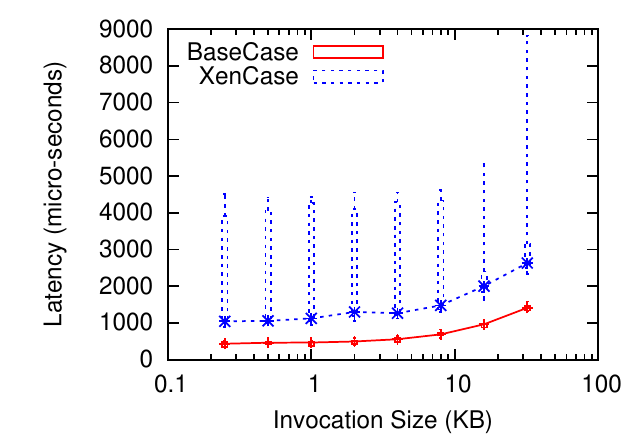,height=2in}\\
                        {(A)}
		\end{tabular}
                \begin{tabular}{c}
                        \psfig{file=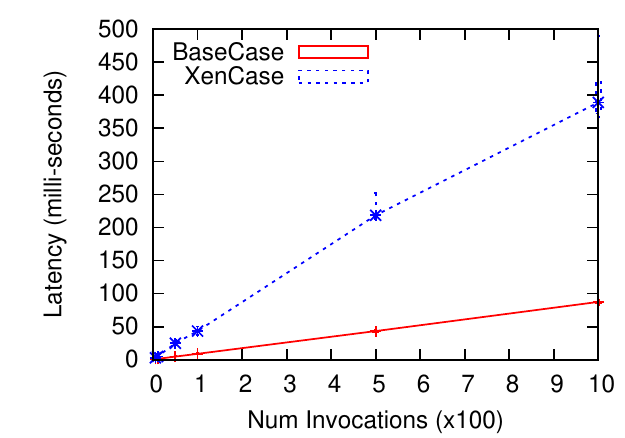,height=2in}\\
                        {(B)} 
                \end{tabular}
        }  
	\vspace{-0.15in}
        \caption{(A) Invocation Costs under different payload sizes
          (of request and response) for two different scenarios (base
          case, xen capsule): Latency (Y-Axis) vs Payload size
          (X-Axis, log scale). The corresponding costs for the TTP
          capsule are:
          $24.4$ ms ($0.25$ KB), $24.7$ ms ($0.5$ KB), $25$ ms ($1$
          KB), $48.1$ ms ($4$ KB), $48.4$ ms ($8$ KB), $72.6$ ms ($16$
          KB), $108.1$ ms ($32$ KB). (B) Invocation Costs for stock
          capsule under two different scenarios (base case, TTP, xen
          capsule): Latency (Y-Axis) vs Number of Ticker Events
          (X-Axis). The corresponding costs for the TTP capsule are:
          $72$ ms ($5$ invocations), $132$ ms ($10$), $612$ ms ($50$),
          $1.2$s ($100$), $6$s ($500$), $12$s ($1000$). }
        \label{fig:invocation_cost}
	\vspace{-0.2in}
\end{figure}

\textbf{Varying Payload Sizes:} To examine the variation of the
invocation cost across different payload sizes, we plotted the latency
as a function of payload size (varied in multiples of $2$ from $256$
bytes to $32$ KB) in Figure~\ref{fig:invocation_cost} (A). In the
plot, the ``cross'' represents the median, the square box represents
the $50\%-$ile, and the lines around the cross represent the
$95\%-$ile.
The plot only shows the base case and Xen case; since the TTP
latencies are much higher, those are listed in the figure's caption.
This plot shows that the overhead added by Xen as a percentage of the
base case declines, while the network transfer time increases the
latency for the TTP case. 
Once again, we note that the variance in the Xen latency is
substantial, especially at low invocation sizes. The bandwidth
consumption for the TTP case varies roughly linear with the size of
the payload, and are $1.7$ KB (for invocation of $256$ bytes size),
$10$ KB ($4$ KB), and $72$ KB ($32$ KB); of course, in the Xen case
and the base case, no network communication is involved for
invocation.


\textbf{Setup Cost:} We now discuss the setup costs in the three
scenarios (which are similar to the tranfer costs from one service to
another) for sensitive data of size $4$ KB. In the base case, the
latency is $13$ ms (primarly due to the $10$ ms round-trip) and the
bandwidth consumed is $6$ KB (the data size of $4$ KB, plus
\textit{libevent} overhead, plus TCP headers and acknowledgement
packets). In both the TTP and Xen case, the capsule code is sent from
the client to the TTP / service, and then invoked from the base
layer. The respective latency figures are $332$ ms and $491$ ms
respectively; we found that this cost was dominated by about $300$ to
load the capsule code into memory and invoke it. We currently use
\textit{dlopen} for this purpose; we intend to replace this when
removing \textit{glibc} from our TCB, so the overhead should be
significantly improved.  The bandwidth cost is roughly the cost of
transferring the capsule code along with the sensitive data; in the
TTP case, this cost is incurred between the user and the TTP, while in
the Xen case, the cost is incurred between the user and the
service. Our main observation here is that, in comparison with the TTP
case, the Xen capsule setup is not significantly higher; the cost of a
two round trip protocol involving key generation for installation is
out-weighed by the cost of the data and code transfer.

\textbf{Summary:} In terms of invocation costs, our main finding is
that, in terms of latency, the capsule case is more expensive than the
base case, but this overhead (as a percentage) decreases with payload
size. In comparison with the TTP case, the latency gain provided by
capsules is significant; even with a low round-trip delay of $10$ ms,
the capsule latencies are significantly better. And of course,
bandwidth consumption proportional of the size of the invocation
request and response are incurred in the TTP case, and not in the
co-location case. In terms of setup costs and storage costs, the main
cost is due to the size of the capsule code itself; we plan to reduce
this in the future by removing the dependence on {\it glibc} and by
altering our compilation tool-chain to produce flat code. However, in
practice, this setup and storage cost may be amortized across several
users who may trust the security company to provide their capsule.

\subsection{Stock capsule} 

\begin{table*}[t]
\begin{scriptsize}
\renewcommand{\arraystretch}{1.15}
\caption{Single Invocation Costs for (A) Stock Capsule (B) CCN Capsule}
\label{tab:stock_ccn_perf1}
\begin{center}
\begin{tabular}{|l|l|l|l|c|}
\hline\hline 
Scenario & \multicolumn{2}{|c|}{Stock Capsule} &
\multicolumn{2}{|c|}{CCN Capsule} \\
\hline
 & Latency (micro-seconds) & Bandwidth (bytes)  & Latency (micro-seconds) & Bandwidth (bytes)  \\
 & (Mode, $50\%$-ile, $95\%$-ile) &  (bytes) & (Mode, $50\%$-ile, $95\%$-ile) &  (bytes) \\
\hline\hline
Base Case & $413~[359,466]~[353,478]$ & NA &
$336~[330.9,344.5]~[322.2,436]$ & NA \\ 
\hline
TTP & $24224~[24143,24303]~[23954,24432]$ & $1294$ & $366.4~[360.2,380.2]~[350.8,550.8]$ & 1327 \\
\hline 
Xen Capsule & $1045~[1001,4179]~[936,4407]$ & NA & $344.6~[336.0,361]~[328.868,523]$ & NA\\ 
 \hline\hline
\end{tabular}
\end{center}
\end{scriptsize}
\vspace{-0.3in}
\end{table*}


Table~\ref{tab:stock_ccn_perf1} shows the invocation cost for the
stock capsule for the three scenarios. The advantage of the Xen option
with respect to the TTP option is clear; co-location offers
significant advantages here since the latency is about $15$ ms lower
and no network bandwidth is consumed. In the financial high-frequency
trading market, this improvement is invaluable; firms typically pay
substantial money to obtain such an advantage. The single invocation
consumes a bandwidth of around $1$ KB for the TTP Case.

In order to reflect a typical stock trading scenario where there are
hundreds of ticker events per section, we plotted the latency metric
versus varying number of back-to-back invocations (corresponding to
ticker events in the stock market) in Figure~\ref{fig:invocation_cost}
(B). Admittedly, this is an apples-to-oranges comparison since the TTP
case requiring network access is substantially slower as compared to
the Xen option and the base case. Comparing the Xen case and the base
case for a sequence of $100$ back-to-back invocations, the latencies
are respectively $43.4$ ms and $9.05$ ms respectively; though the
overhead due to Xen is substantial, it is still a significant
improvement than the TTP case which requires $12$ seconds. Regards the
bandwidth, in the TTP case, the bandwidth consumed for a sequence of
$5, 10, 50, 100, 500$ and $1000$ invocations are respectively $2.8,
4.8, 20.2, 39.4, 194, 387$ KB respectively. In our experiments, we
have only considered one symbol being traded; typically, a user would
be interested in trading on tens or hundreds of different symbols,
which would lead to a proportional increase in the bandwidth. Thus,
the bandwidth required for say, $500$ events per second, is about
$1.6$ MBps per user. Though well within the ability of today's
network, this bandwidth still to be provisioned and paid for, and can
be expensive.

These results reflect that the flexibility of the framework with
respect to deployment at the service site is a significant benefit in
terms of network bandwidth. The capsule option is beneficial to both
the stock broker and the user; the stock broker need only invest in a
trusted module on its computers and avoid paying for bandwidth, while
the user obtains the advantage of housing her trading strategy very
close to the data source itself thus avoiding wide-area network round
trips.

\subsection{CCN capsule}


We present the latency and the bandwidth consumption of a single
invocation of the capsule in Figure~\ref{tab:stock_ccn_perf1}. The
invocation here involves interacting over the network with the payment
gateway over the wide-area over SSL, and thus the additional overhead
in the Xen Capsule or the TTP case is minimal as compared with the
base case. The Xen capsule, in this case, consumes $10$ ms more as
compared to the the base case, and is in fact, more expensive than
even the TTP case. This is due to the fact that we have offloaded the
network stack to the host-hub; the SSL negotiation before the HTTP
request is sent requires many packets to be sent, which involves
several context switches between the two VMs, which leads to the
increase in latency. We plan to reduce this overhead in the future by
implementing a front end and back end specific to SSL which would farm
out the significant portion of the computation to the host hub; the
SSL front end in the capsule would only be responsible for (a)
certificate verification (b) session key generation. This would help
us both reduce the overhead as well decrease the TCB size. The credit
charging request itself is a one-step HTTP request response protocol;
thus this latency can be reduced by pre-establishing SSL
communications or re-using session keys to avoid SSL negotiation
during invocation. The latency numbers also reflect some outliers in
the $95\%$-ile range; this is due to delay outliers at the payment
gateway itself, and are not related to our framework itself. A nominal
bandwidth of about $1.3$ KB is consumed in the TTP case. Once again,
we emphasize that these results demonstrate the cost of the generality
in our framework as compared to custom solutions already deployed
today; the main gain in the CCN case is due to the flexibility of
policy, not performance.

\subsection{Targeted Ads capsule}

\begin{figure}
	\vspace{-0.1in}
        \centering
        \mbox{
                \begin{tabular}{c}
                        \psfig{file=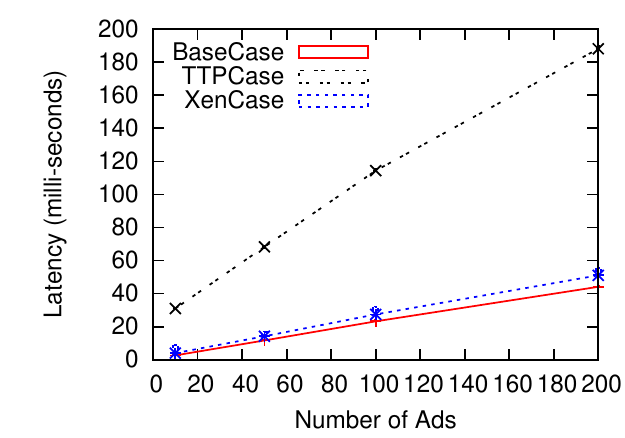,height=2in}\\
                        {(A)}
		\end{tabular}
\begin{tabular}{|l|l|l|}
\hline\hline 
Scenario & Base/Xen Case & TTP \\
\hline\hline
2 & 46.5 & 62.1 \\
\hline 
5 & 139.6 & 201.8 \\ 
\hline
10 & 294.7 & 434.7 \\
\hline
25 & 715.2 & 1133.3 \\
\hline
50 & 1535.9 & 2297.2 \\
\hline
100 & 3087.2 & 4626.5 \\
\hline\hline
\multicolumn{3}{|c|}{(B)} \\
\hline
\end{tabular}
        }  
	\vspace{-0.15in}
        \caption{ (A) Invocation Costs for targeted ads capsules that
          serves ads under three different scenarios (base case, TTP,
          xen capsule) for the \textit{SelectAd} interface: Latency
          (Y-Axis) vs (Number of Ads) (X-Axis). (B) Aggregation
          Bandwidth Requirement: Column titles denote number of users,
          Row titles reflect the scenario, and the matrix element
          denotes the bandwidth consumed (KB). Individual user
          histories are about $15$ KB per user. }
        \label{fig:ads_perf_aggreg}
	\vspace{-0.2in}
\end{figure}



 
We present two sets of results for the targeted ads scenario.
Figure~\ref{fig:ads_perf_aggreg} (A) shows the latency per invocation
for the capsule to serve targeted ads using the {\it SelectAd} (we do
not show results for the {\it GetInterestVector} interface due to lack
of space). This graph shows that the Xen capsule has nearly the same
overhead as the Base case capsule once the number of ads out of which
the capule selects one exceeds $5$. This is mainly because the payload
size dominates the transfer time. In the TTP case, the latency is
clearly impacted by the wide-area network delay. The bandwidth
consumed in the TTP case was measured to be $2.6, 8.3, 15.8, 29.2$ KB
for $10, 50, 100, 200$ ads respectively. These reflect the bandwidth
savings of the co-located capsule option; since this bandwidth is
incurred for every website the user visits, this could be significant.


There are two other operations supported by this capsule which are of
interest: allowing a user to update her web history and the data
aggregation transformation. For these operations, we present the
bandwidth consumed, since latency is not the primary constraint in
these cases. Regarding the update operation, the bandwidth required
was measured to be $2.1, 4.8, 15.9, 23.3, 52.7$ KB for $5, 10, 50,
100, 200$ new websites visited per day by the user. The average number
of websites visited per user is typically around $50$ websites per
day~\cite{web:acmweb08}, so we chose to experiment in the range of
$5-200$ websites per day. This bandwidth is incurred in the
server-side capsule, the base case, and the TTP case; it however is
not necessary in the client-side capsule. In the case of targeted
advertising, we expect that the update traffic per day is much lower
than the cost of sending ads (which change much more frequently); thus
the server-side co-located capsule is advantageous with respect to
bandwidth consumption. In other cases where user data is much more
frequently updated, a client-side capsule may be more suitable.

Regarding the aggregation operation, we currently employ an
aggregation strategy that optimizes the total bandwidth consumed. In
these experiments, we consider $N$ users who have decided to pool
their data together. In the base case and Xen case, we assume that the
data of each user is hosted on a physically different machines in the
data-center; the optimal way to combine their data in terms of
bandwidth is for a simple hub-and-spoke model. This is accomplished by
the local host-hub for users $1,\cdots,(N-1)$ executing a {\it pull}
operation on their capsules, and pushing the resultant encrypted
histories to the host-hub for user $N$. At this point, the capsule for
user $N$ would release the aggregate data upon request. In the
TTP/client-side case, we assume that the $N$ users have each entrusted
their data to $N$ different administrative entities; in this case, the
bandwidth-optimal strategy is for the service is a similar
hub-and-spoke strategy. We experimented with varying numbers of users
and present the bandwidth consumed in Figure~\ref{fig:ads_perf_aggreg}
(B). In this experiment, the data per user is fixed to be $1K$ per
user. The bandwidth consumed for the Xen case and the base case are
the same since the same strategy is involved; the TTP case consumes
much more bandwidth since both the push and pull requests require
remote access, as does the final pull from capsule $N$. Thus, the
advantage in the Xen case is that the bandwidth is lower, and further
the bandwidth consumed is only intra-data-center. Further, the
bandwidth consumed can be optimized much more easily for the
server-side capsule and the base case since it is within the confines
of a single administrator. This demonstrates the usefulness of the
aggregation operation especially when thousands of users are involved.




\section{Security Analysis} \label{sec:secanalysis} 

We now discuss the desirable security properties of our capsule,
determine the TCB for these security properties to hold, and argue
that our TCB is a significant improvement over the status quo.

\subsection{Security Properties}

\textbf{Security Goal:} Our over-arching security goal (from our
problem statement in Section~\ref{sec:problem}) is that any access to
the data or any transfer of the data must only occur in accordance
with the interface and policies sanctioned by the user. The adversary
is a party that has compromised the software stack at the service or
is a party that has physical access to the machine hosting the
capsule. Note that this goal offers protection only to the extent to
which the data is protected by the interface and user's policies. For
instance, if the user specified a budget of $100$ dollars for her
credit card, and the machine hosting her capsule was compromised, her
card could be charged upto her budget by the compromise. However, such
a charge would still be securely tied to the compromised service since
it is not possible to spoof the invoking principal, thus offering
sufficient data for post-mortem forensics. Further, we note that the
service can always launch an availability attack in the co-located
case by simply refusing to service requests using its host hub or by
denying access to a user in accessing her capsule to, say, retrieve a
record of all operations; we only aim to prevent the adversary from
violating the policy constraints, not liveness requirements.

\textbf{Informal Security Argument:} We will now present informal
arguments as to why our capsule framework acheive this security goal.
In the case of a TTP-based deployment, the argument is simple; we rely
on the isolation provided by the TTP and the correctness of the
capsule implementation to ensure the security goal. The rest of this
section is concerned with the co-hosting case. We are currently
exploring using the $LS^{2}$ framework~\cite{ls2:oakland09} to state
and formally verify our security goals in the co-hosting scenario; we
defer such any formal argument for future work. In the following
informal arguments, we will assume that the trusted module provides
the four security properties detailed in
Section~\ref{sec:design_interaction}: isolation, remote attestation,
anti-replay protection, and random number generation.

We make this argument in two steps. First, we argue that any
interaction between the service and the user's data occurs only via
the functions exported by the capsule base layer. Second, we will rely
on the base layer protocols and its correctness to ensure that all
these interactions are in conformance with the user's policies.

For the first, we will rely on the isolation property of the trusted
module. In the case of the VM, this assumes that the VM hosting the
service code can interact with the capsule VM only via the XenSocket
interface. This requires that the VMM be placed in our TCB. We also
assume that the VMM prevents significant leakage of information
through side-channels such as memory page caching, CPU usage: this is
a subject of ongoing research (\eg
Tromer~\etal~\cite{arch_attacks_mitigiation:sosp09}). For the second,
we note that the following lists all the possible interactions between
the service and the capsule: installation, invocation, hosting
transfer, and data transformation (going through the parts of the
framework in Section~\ref{sec:arch}). We will now consider each in
turn.

\textbf{Installation:} The installation protocol
(Section~\ref{sec:design_hosting}) provides three properties in this
regard. First, it uses the attestation functionality of the trusted
module to ensure that only the base layer code generates the public
key in response to the first installation message. Second, the
installation protocol (based on the classic Diffie-Hellman protocol
with RSA keys exchange alongside the Diffie-Hellman messages) ensures
the confidentiality of the information; only the party which generates
the public key in response to the first installation message can
decrpt the capsule. Due to the attestation guarantee, this party is
guaranteed to the base layer, which is trusted. Note we rely on the
random number generation ability of the trusted module to ensure that
the generated public key is truly random and cannot be biased or
guessed by the adversary. Third, since the attestation verifies the
purported input and output to the code which generates the public key,
man-in-the-middle attacks are ruled out; an adversary cannot alter any
part of the message without rendering the atttestation invalid. These
three properties ensure in conjunction that the capsule can only be
decrypted by the base layer. We then rely on the correctness of the
base layer to argue that the capsule is correctly decrypted and
instantiated at the service.

\textbf{Invocation:} We next need to argue that any invocation occurs
in accordance with user's policies. Every invocation is checked by the
base layer to ensure: (a) the identity of the invoking principal (b)
the correctness of any supporting policies (c) that the user-specified
policies along with the supporting policies allow the invoking
principal the privilege to make the particular invocation.

The identity of the invoking principal is verified either against a
list of public keys sent during the installation protocol (this data
list binds the names of services to their public keys) or, in the case
of the owner, it is verified against the public key of the data owner
sent during the installation protocol. In either case, the correctness
of the installation protocol ensures that the identity cannot be
spoofed since the list of public keys used to validate the identity is
correct. The correctness of the supporting policies is ensured by
verifying that each statement of the form $X~\underline{says}~\cdots$
is verified against the signature of $X$. The policy resolver takes
two sets of policies as input: the database of policies sent by the
user and the set of supporting policies sent by the invoker. The
correctness of the latter we have already argued for; for the former,
we rely on the anti-replay property provided by the trusted module to
ensure that the database of policies in the capsule (which can be
updated over time) is up-to-date and reflects the outcome of all the
invocations made so far. We assume the correctness of the policy
resolver to ensure that it answers invocation queries correctly in
accordance with these policies. For stateful constraints, the
implementation also updates the state before returning the result of
the invocation to the service.

Once the invocation is granted by the base layer, it is passed on to
the data layer which carries out the required functionality. In cases
such as a query capsule or an analytics capsule, this functionality is
carried out entirely within the data layer, which we assume to be
correct. In cases such as the proxy capsule, which requires access to
the network, we note that though the network stack is itself offloaded
to the host hub, we rely on the host hub only for availability, not
for safety. The SSL library and cryptographic library are enclosed
within the capsule, and thus, it does not matter that the host hub is
effectively a man-in-the-middle. In this respect, we also note that
the other services provided by the host hub, namely, persistent
storage and obtaining the current time, are similarly relied only for
availability, not safety, since the base layer validates persistent
storage using its own integrity and confidentiality checks (once
again, relying on the anti-replay functionality provided by the
trusted module), and verifies the current time by validating the
source (such as a trusted NTP server).

We next consider the hosting protocol. Since it consists of the
invocation protocol (used by the service to make the transfer request)
and the installation protocol (used to effect the transfer), we rely
on our arguments regarding invocation and installation to argue that
hosting is in accordance with our security goal as well.

\textbf{Data Transformation:} We now turn our attention to the data
transformation protocol. The filtering and update protocol happen only
within the confines of a single capsule, and their correctness
arguments is similar to those for invocation. The aggregation protocol
though more complicated decomposes into a number of simple pair-wise
aggregation operation. Since during the pair-wise operation, a capsule
verifies the identity of its peer as belonging to a user who is part
of the data crowd, the data is merged only between capsules that are
part of the same data crowd. Since the merged information is only
revealed to the service after a minimum number of capsules belonging
to the data crowd are merged, this aggregation is in accordance with
the user's policies.

\subsection{TCB} \label{sec:secanalysis_tcb}

We now evaluate the TCB of our capsule framework. From the previous
section, it is clear that our TCB consists of: (A) the trusted module
(B) the data layer interface and implementation (C) the base layer
protocols and implementation.

The trust assumption in (A) appears reasonable for VMMs, TPMs, and a
secure co-processor. A carefully managed VMM which presents a minimal
attack surface to applications inside a guest VM may be judged to be
adequate enough to defend against software threats. Similarly, TPMs
and secure co-processor have both been designed across several
iterations; it seems reasonable to expect that their designs and
implementation in today's chipsets will be difficult to exploit.
Regarding (B) and (C), the LOC metrics are discussed in
Section~\ref{sec:impl_framework}). We base our trust in the data layer
interface and implementation in the simplicity of the interface; the
data specific portion of the interface is less than $500$ lines even
for our most complicated capsule (the CCN capsule). For other capsules
following the query idiom, analytics idiom, and provenance idiom, the
implementation is even simpler since no network access is required.

Currently, the base layer protocols and implementation are much more
complex than the data layer itself, as reflected in the LOC metric
(the base layer has roughly 6K LOC). While we have aimed to keep our
design simple right from a minimal policy language to offloading
network access to the untrusted host hub, the base layer is still
complex, and currently includes a cryptographic library (PolarSSL) as
well. In the future, we hope to formally verify the correctness of
this implementation. For now, our only argument is that even granting
that the base layer is complex, it offers a significant improvement in
security to users of web services today who have no choice but to rely
on an unaudited closed source service code for their privacy
guarantee. More importantly, it restores control to the user over her
data, allowing a security conscious user to invest in a secure capsule
implementation for her data from a trusted security company. Another
comparison point in this respect is an information flow control system
(such as JIF or Asbestos). The TCB in such a case includes the
run-time (in Jif's case, this is the Java run-time along with Jif's
typing rules; in Asbestos's case, it would be the operating system)
along with any trusted de-classifiers (a declassifiers is a code block
or a process that is allowed to break the typing rules or propagation
rules; such declassifiers are essential to prevent over-conservative
information flow propagation). We believe this compares favorably with
our TCB; building a general enforcement mechanism seems a more complex
undertaking than implementing a simple interface which makes our
approach easier to verify.

\section{Related Work} \label{sec:related}

We first discuss two closely related papers in literature, before
examining broader related work. The closest related work to our own is
in Wilhelm's thesis~\cite{mobile_privacy:thesis} (Chapter 5). This
work leverages mobile agents (a well-investigated research paradigm;
an agent is code passed around automatically from one system to
another in order to accomplish some functionality) to address the
issue of privacy. The suggestion is to encapsulate data with a generic
policy enforcement layer that provides a fixed set of functionality
(such as, enforce time for retention, audit operations). The main
differences from our approach are: (a) our work is tailored towards
web services; in particular, operations like hosting transfer and data
transformation were chosen to model how web services handle data today
(b) our approach is extensible to various kinds of sensitive data; the
enforcement layer is configurable by the user and is specific to the
particular kind of sensitive data (c) we aim to design the capsule
interface itself so as to reduce data exposure. The second closely
related paper to ours is by Iliev~\etal~\cite{client_privacy:secpriv}.
They propose the use of trusted hardware to offer client privacy when
operating on server data for two applications: private information
retrieval (without traditional cryptographic mechanisms) and an
armored network traffic vault. This proposal is similar to our
co-location option. Two characteristics of our framework that are
pertinent to web services, and do not apply to
\cite{client_privacy:secpriv} are: (a) support for scenarios where a
user's data is spread across multiple services. (b) policy-based
control over invocation, hosting, and data transformation. Apart from
these specific papers, our work is related to the following broad
areas:


\noindent \textbf{Information Flow Control (IFC):} The principle of
IFC has been implemented in OSs (\eg Asbestos~\cite{asbestos:sosp05})
and programming languages (\eg JIF~\cite{jif:jsac03}), and allows the
control of flow of information between multiple processes or security
compartments. IFC has also been used to build secure frameworks for
web services in W5~\cite{w5:hotnets07}, xBook~\cite{xbook:ssym09}. The
main difference from the capsule framework is that we provide data
access control, not data propagation control. Privacy is guaranteed by
the interface chosen by the user, and not by run-time policy
enforcement. Capsules apply only to cases where an interface can be
arrived at that offers sufficient privacy to the user as well as is
usable to the service; if no such interfaces exist, we can leverage
IFC frameworks to control the propagation of sensitive
information. The advantage in restricting to interface-based access
control is that we can rely on a variety of isolation mechanisms (such
as TPMs) without requiring a particular OS or programming
language. Further, the simplicity of the interface makes it feasible
to envision the possibility of proving the correctness of capsule
code; doing so in the IFC case requires one to prove the correctness
of the enforcement mechanism (OS or compiler) which can be
significantly more complex.

\noindent \textbf{Decentralized Frameworks For Web Services:} Privacy
frameworks that require only participation from users have been
proposed as an alternative to web services. VIS~\cite{vis:mobiheld09}
maintains a social network is maintained in a completely decentralized
fashion by users hosting their data on trusted parties of their own
choice; there is no centralized web service. Capsules are more
compatible with the current ecosystem of a web service storing user's
data and rely on the use of interfaces to guarantee privacy.
NOYB~\cite{noyb:wosp08} and LockR~\cite{lockr:conext09} are two recent
proposals that rely on end-to-end encryption to hide data from social
networks; both these approaches are specific to social networks, and
their mechanisms can be incorporated in the capsule framework as well,
if so desired.

\noindent \textbf{Enterprise Privacy Architectures:} Several privacy
architectures that deal with the propagation of information within an
enterprise or across closely-related enterprises have been proposed.
These include Ashley~\etal~\cite{platform_enterprise:isse02},
Backes~\etal~\cite{toolkit_privacy:esorics03},
GeoDac~\cite{geodac:tech_rep}. Broadly, these papers propose
specification languages and enforcement mechanisms that help enforce
enterprise-wide policies on data retention, access logging, and other
accountability guarantees. The main difference from the capsule
framework is that (a) our framework involves two mutually distrusting
parties, the user and the service, and thus we only base enforcement
on hardware or system software, rather than application software. (b)
our framework is designed for web services. (c) our goal is to enforce
interface constraints and policy control that control data leaks,
rather than to enforce generic data control policies. In future, we
hope to incorporate guarantees provided by these architectures in our
framework as well. Our provenance capsule is also related to
VFIT~\cite{vfit:iacs08} and GARM~\cite{garm:hotsec09} which follow the
modification of data across multiple machines.

\noindent \textbf{Privacy in Cloud computing:} The area of cloud
computing has seen a lot of work in the context of privacy as well.
These include Trusted Cloud~\cite{trusted_cloud:ssym09}, Accountable
Cloud~\cite{accountable_cloud:ladis09}, Cloud
Provenance~\cite{provenance_cloud:ladis09}. These works deal with the
more complex problem of guaranteeing correctness of code execution on
an untrusted third party. Capsules are only concerned to protecting
the privacy of the user's data; we assume the application service
carries out the service (such as sending correct ticker data) as
expected. Airavat~\cite{airavat:nsdi10} proposes a privacy-preserving
version of MapReduce based on information flow control and
differential privacy; capsules support general kinds of computation,
but have to rely on manual re-factoring whereas Airavat can
automatically ensure privacy by restricting itself to specific types
of MapReduce computations. We also note that attacks based on leakage
across VMs are known~\cite{get_off:ccs09} and defense mechanisms
against such attacks are also being
developed~~\cite{arch_attacks_mitigiation:sosp09}. Our capsule
framework can avail of such defense mechanisms as they are developed
further; we view such work as orthongal to our central goal.

\noindent \textbf{Targeted Advertising Systems:}
PrivAD~\cite{privad:hotnets09} and Adnostic~\cite{adnostic:ndss10} are
recent proposals that are client-side systems for targeted
advertising; this is somewhat similar to a client-side targeted ads
capsule in our framework. The difference is that our framework
generalizes to other deployment scenarios as well. In the future, we
hope to borrow their techniques for anonymized ad impression
collection in our targeted ads capsule as well; currently, we do this
using aggregation, PrivAD and Adnostic offer dealer-based and
encryption-based mechanisms respectively for this purpose.

\noindent \textbf{Mechanisms:} The capsule framework builds on
existing isolation mechanisms, such as the virtual machine security
architecture (\eg Terra~\cite{terra:sigops03}), proposals that use
TPMs (\eg Flicker~\cite{flicker:eurosys08}), and systems based on a
secure co-processor~\cite{yee:thesis}. These proposals offer the
foundation upon which our capsule framework can build on to provide
useful guarantees for data owners. Our implementation also borrows
existing mechanisms (XenSocket~\cite{xensocket:middle07},
vTPM~\cite{vtpm:ssym06},
disaggregation~\cite{xendisaggregation:sigops08}, use of hardware
virtualization features~\cite{secureinvm:ccs09}) that help us improve
the performance and security of a virtual machine based architecture.
Currently, we do not provide any automatic re-factoring mechanisms; in
the future, we hope to explore using existing program partioning
approaches (\eg Swift~\cite{swift:sosp07},
PrivTrans~\cite{privtrans:ssym04}) for this purpose.

\section{Conclusion and Future Work} \label{sec:conclusion}

Our capsule framework applies the well-known principle of
encapsulation to provide data access control guarantees to users who
wish to avail of web services that operate on their data whilst still
retaining control over the use of said data. Our design goals are to
allow flexible deployment of such capsules and to provide flexible
policy control to their users. The capsule approach works when a
satisfactory interface can be arrived at that gives privacy to the
user while still sufficing to carry out the service: we show this to
be the case for a broad variety of currently operating web
services. For others, one can rely on information flow control
techniques or end-to-end encryption to obtain privacy guarantees. Our
prototype implementation includes four kinds of capsules as examples;
although the performance of the co-located Xen-based capsule is not
quite suitable for deployment in a production service, they do offer
substantial advantages over the TTP and client-side capsules in the
dimensions of network bandwidth consumption and/or provisioning cost.

We identify three fruitful avenues for future work. First, our
implementation can be improved by incorporating known virtualization
techniques to reduce overhead and by removing the need for including
system libraries and kernels in our TCB by a more careful
implementation. Second, formal verification of the control layer
implementation and its security protocols would allow for much greater
confidence in its correctness as compared to today. We have done some
preliminary work in this area towards expressing our implementation in
a formal language and proving that it conforms to our security
expectations. 
Similar correctness guarantees would be useful for the data layer
interface and implementation as well. The flavor of correctness proofs
here is different from those in the control layer protocol; the goal
is to prove that there is no leakage of information via the
implementation. One can envision static information flow checking
(such as JIF~\cite{jif:jsac03}) to provide such a guarantee, or a
proof from scratch based on computational non-inference. Third, given
the complementary nature of information flow control systems, it would
be useful to integrate such elements into the capsule framework;
interface calls which do not leak any information would be part of the
capsule framework, leaky interface calls can leverage use of the
information flow policies. Of course, the former would involve a
smaller TCB, since information flow requires placing the information
flow enforcement kernel
in the TCB.\\

\noindent \textbf{Acknowledgements:} We thank the authors of Adnostic,
in particular, Arvind Narayanan, for sharing their data.

\bibliographystyle{ieeetr}
\bibliography{conf}

\end{document}